\documentclass[
 authoryear]{FLO_v1}%

\usepackage{amsfonts,amsmath,amssymb,amsthm}
\usepackage{appendix}
\usepackage[T1]{fontenc}
\usepackage{graphicx}
 \graphicspath{{./figs/}}
\usepackage{ifpdf}
\usepackage{mathrsfs}
\usepackage{multicol,multirow}
\usepackage[authoryear]{natbib}
\usepackage{newtxmath}
\usepackage{newtxtext}
\usepackage[figuresright]{rotating}
\usepackage{textcomp}
\usepackage{upgreek}
\usepackage{xcolor}
\usepackage{varwidth}
\usepackage[shortlabels]{enumitem}

\usepackage{atveryend}
\usepackage{mismath}
 \pinumber[uppi]
\usepackage{verbatim}

\usepackage[colorlinks,allcolors=blue]{hyperref}
\definecolor{jourcolor}{cmyk}{1,0.57,0.01,0.38}
\hypersetup{
    colorlinks,%
    citecolor=jourcolor,%
    filecolor=jourcolor,%
    linkcolor=jourcolor,%
    urlcolor=jourcolor
}

\theoremstyle{definition}

\newcommand\bcdot{\boldsymbol{\cdot}}
\newcommand\Fr{\mbox{\textit{Fr}}} 
\newcommand\Pra{\mbox{\textit{Pr}}} 
\newcommand\Rey{\mbox{\textit{Re}}} 

\articletype{RESEARCH ARTICLE}

\DOI{10.1017/flo.2023.xxx}

\Year{2023}

\Vol{x}

\Price{}

\art-id{FLO2000049}

\citearticle{Wagner, L., Braun, S. \& Scheichl, B.}

\begin{document}

\title[Opening a champagne bottle]
{Simulating the opening of a champagne bottle}

\author[Lukas Wagner, Stefan Braun and Bernhard Scheichl]
{Lukas Wagner$^{1,2*}$\! {\href{https://orcid.org/0009-0006-9110-1430}{\includegraphics[width=0.02\textwidth]{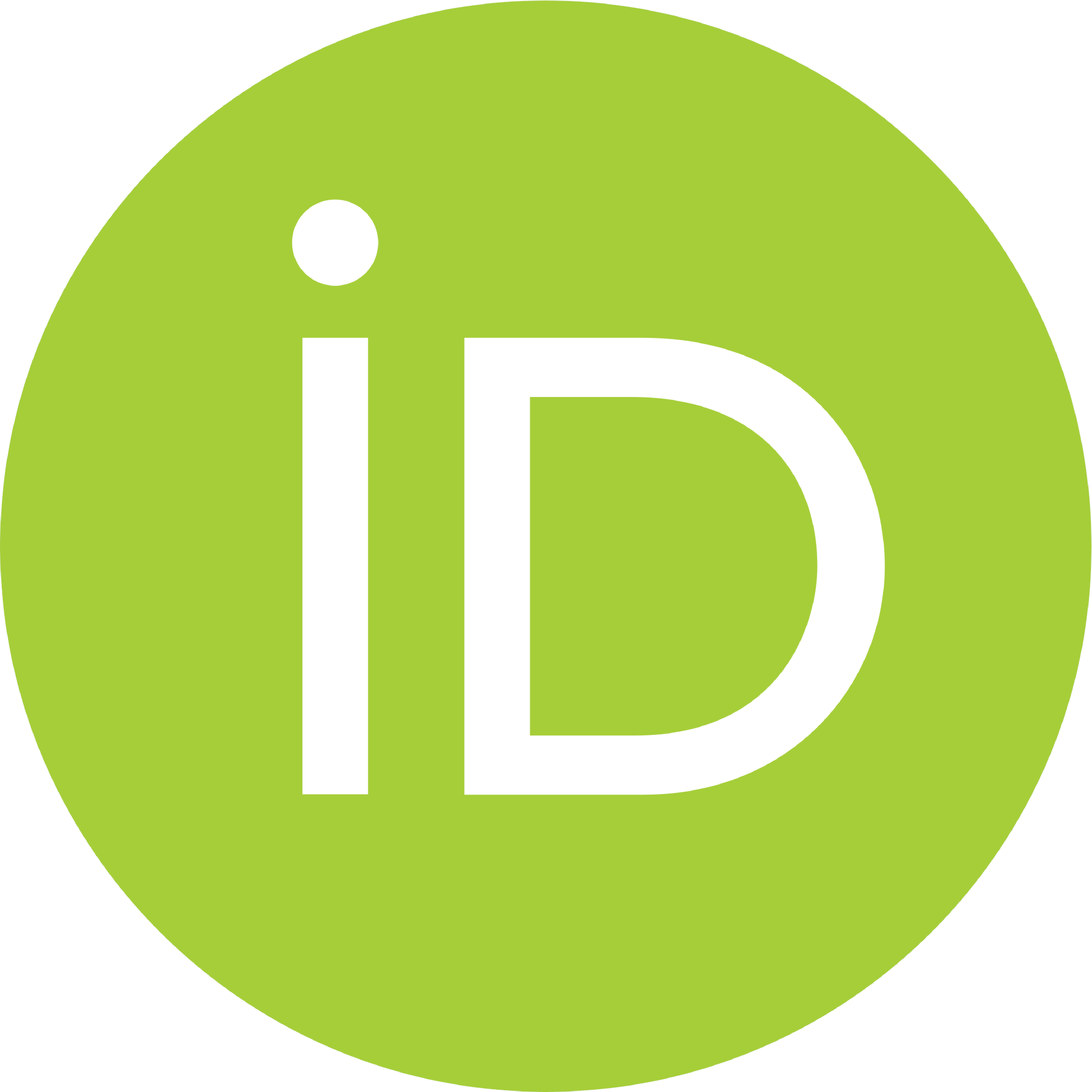}}}
Stefan Braun$^{1}$\!
{\href{https://orcid.org/0000-0002-7145-1103}{\includegraphics[width=0.02\textwidth]{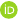}}}
and Bernhard Scheichl$^{1,2}$\!
{\href{https://orcid.org/0000-0002-5685-9653}{\includegraphics[width=0.02\textwidth]{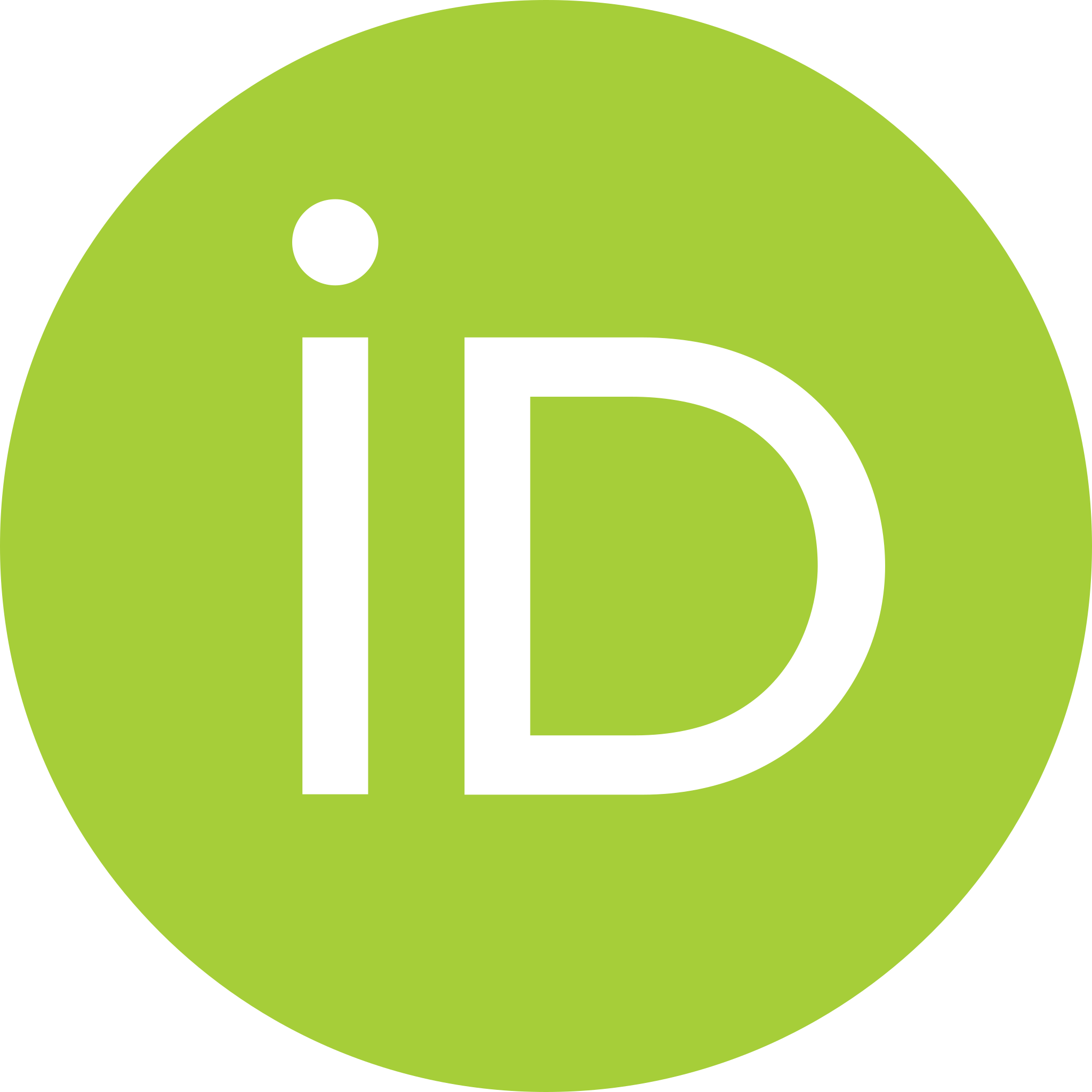}}}}

\address[1]{Institute of Fluid Mechanics and Heat Transfer, TU Wien, Tower BA/E322, Getreidemarkt 9, 1060 Wien, Austria}
\address[2]{AC2T research GmbH (Austrian Excellence Center for Tribology), Viktor-Kaplan-Straße 2/C, 2700 Wiener Neustadt, Austria}
\corres{*}{Corresponding author. E-mail: \emaillink{lukas.wagner@ac2t.at}}

\keywords{Discharge flows, Fluid--Structure interactions, Jet formation, Mach disc, Shock waves, Transitional ballistics}

\date{\textbf{Received:} 28 04 2023; \textbf{Revised:} 15 09 2023; \textbf{Accepted:} 17 10 2023}

\abstract{The axially symmetric, swirl-free gas dynamics and interlinked motion of a cork stopper provoked by the opening of a champagne bottle are modelled rigorously and studied numerically. The experimental study by Liger-Belair, Cordier \& Georges (\textit{Science Advances}, \textit{5}(9), 2019) animated the present investigation. Inspection analysis justifies the inviscid treatment of the expanding jet of air enriched with dissolved carbonic acid gas initially pressurised in the bottle. Solving the resulting Euler equations is facilitated through the open-source software \textit{Clawpack}. Specific enhancements allow for resolving the emerging supersonic pockets, associated with surprisingly complex shock structures, as well as the gas--stopper interaction with due accuracy. Our experimental effort provided modelling the frictional behaviour, constitutive law and reversible (de-)compression of cork. Initially, the gas expands inside the bottleneck yet sealed by the stopper, hence accelerated by the gas but decelerated by dry sliding friction. Once the stopper has passed the bottle opening, the jet rapidly assumes locally supersonic speed, where a complex shock pattern is detected. Special attention is paid to the formation and dissolution of one or even two Mach discs between the opening and the released stopper. These simulated dynamics are found to be in fairly good agreement with recent experimental findings.}

\maketitle

\begin{boxtext}
\textbf{\mathversion{bold}Impact Statement}
Popping of the cork stopper sealing a champagne bottle unequivocally represents an appealing prototype of a daily-life event that offers a remarkably rich variety of physical phenomena at play. Here these accompany the sudden opening of, in more general terms, a vessel partially filled with a pressurised liquid and dissolved gas (mostly carbon dioxide). Extending the detailed numerical investigation put forward to the situation of more complex geometries and/or different sealing devices having different material properties is an indicated task of further activities. The current focus on the involved gas dynamics, highly unsteady and supersonic albeit at relatively moderate Mach numbers, might already provide a better insight into the complex details, specifically the shock structures, of transitional ballistics in related but more extreme situations of engineering importance. The study is also viewed as a benchmark test for future experimental and numerical efforts, in particular the spatial/temporal resolution of the full fluid--structure interaction of a high-speed gas flow with a solid moving obstacle. We suggest further modelling activities to predict (amongst others) the impact of temperature variations on the initial pressure inside the bottle by virtue of the solubility of carbon dioxide and the limitations of the inviscid-flow model for such scenarios, given the subtle role of viscosity lurking.
\end{boxtext}

\section{Motivation and introduction}\label{s:mi}

The quite recent experimental study by Liger-Belair, Cordier \& Georges (2019), succeeding the pioneering ones by \cite{VoMoe12} and \cite{Lietal13,Lietal17}, discloses the surprisingly complex formation of an expanding gas jet, propelling the cork stopper out of a just opened champagne bottle where they generate the typical popping sound by the radiation of shock waves. This stimulated us to tackle the challenge of resolving this process numerically in due detail, notably by resorting to the forerunner investigation by one of us \citep{Wa21}. It is the ultimate goal of this work to supplement the experimental findings with theoretical predictions based on high-resolution computations. To this end, the modelling focus is laid on the essential physical mechanisms while effects of subordinate importance, either at play on smaller temporal/spatial scales or such as spontaneous phase transition, are deliberately ignored.

Taking up a rigorous point of view sets the basis for a least-degenerate, sound mathematical description of the problem that combines the fields of gas dynamics, fluid--structure interaction and solid--solid friction. Simultaneously, viscous effects are found as insignificantly weak over the spatial and temporal gross resolution of the jet propagation. 
While the compressed gas inside the bottleneck technically contains CO$_2$ gas and water vapor besides small traces of alcohol, the first partial pressure vastly outnumbers the other ones and is therefore considered as a single-phase with a heat capacity ratio at atmospheric pressure and $T=-40$°C of $\kappa_{\mathrm{CO}_2}\approx1.337$ \citep{NIST23}. The ambient air surrounding the bottle has a ratio of $\kappa_\mathrm{air}\approx1.401$ \citep{ETB}. The fact that the temperature of the system can fall below $-100$°C and $\kappa_{\mathrm{CO}_2}$ increases faster with decreasing $T$ than $\kappa_\mathrm{air}$ does, explains why the difference between their ratios in the region of the Mach disc is even smaller. Additionally, the total mass of air vastly exceeds the one of CO$_2$, explaining the nearly instantaneous mixing of both gases. Both considerations allow the assumption of a single-phase gas with a constant heat capacity ratio of $\kappa=1.4$ obeying the ideal gas law.
%
This alleviates the computational effort drastically while the (viscosity-affected) mixing process itself would not alter the results substantially. The fraction of time within which the process is simulated spans from the effective opening of the bottle, seen as instantaneously fast, to the decay of the Mach disc forming just outside of the bottleneck when the jet-driven stopper is already sufficiently remote from the latter.

We concede that the entire neglect of both viscous forces and the co-existence and mixing of two gaseous phases appears doubtful in variety of possible scenarios. To name the most critical ones, these include the shearing flow through the very small gap as the stopper just passes the bottle opening, gross separation of boundary layers accompanied by vortex formation as well as partial freezing of the carbon dioxide (CO$_2$) during its sufficiently strong (isentropic) expansion, observed by \cite{Lietal19}. However, these simplifications are substantiated even though by dimensional reasoning based upon the geometrical and kinematic reference scales. Consequently, they are admissible as their local/sudden invalidity would not affect severely the simulated dynamics of the stopper and the shock structure elucidated on those scales and being at the centre of our attention.

The resulting inviscid-flow simulations are preferably built upon the well-established open-source software environment \textit{Clawpack} \citep{CP22,Maetal16,Le02}, centred around the usual combination of Godunov's numerical finite-volume scheme and Roe's approximate Riemann solver for hyperbolic systems of equations. Special care is taken to facilitate the complete gas--stopper interaction in a fully implicit manner and the stage of stopper--bottleneck interaction initiating the release of the stopper. Therefore, our approach contrasts with the recent computational study by \cite{Beetal22}, complementing the aforementioned motivating experimental work: these authors employ the widely used commercial flow simulation software package \textit{Ansys Fluent}\textsuperscript{\textregistered} to solve the fully Reynolds-averaged Navier--Stokes equations using turbulence closures but disregard that initial stage entirely. The present investigation, however, aims at resolving the coupled gas and stopper dynamics of the whole process in greatest possible detail where our focus lies on the formation of the Mach disc. It might thereby set the basis for a discussion and, in turn, proper inclusion of the here neglected viscous effects in this and closely related situations. In order to establish a reliable model of its compression/expansion and friction, tribological tests with a cork potentially in use were carried out at AC2T research GmbH.

The paper is organised as follows. At first, we argue the simplifications addressed above to set up the corresponding physical-mathematical model (\S\,\ref{s:pmp}). This comprises the description of the gas and air flow, its interaction with the propelled stopper, subject to the dry sliding friction with the bottle prior to its expansion outside of this. We then outline in due detail the incorporation of particular numerical strategies in \textit{Clawpack} that succeed in the stable resolution of the fully coupled gas--stopper interaction and solid--solid sliding (\S\,\ref{s:smp}). As the bedrock of our study, the discussion of the results (\S\,\ref{s:rpd}) is accompanied by their comparison with those available in the literature and addressed above (\S\,\ref{ss:psc}). Finally, an outlook towards potential further implications completes the study (\S\,\ref{s:sfo}).

The Supplementary Material contains two simulation movies and additional information, primarily addressing the technically interested reader, as ``Supplementary technical details'' (comprising the Supplements A--D). However, the paper is self-contained to the necessary extent without the latter.

\section{Physical model and problem formulation}\label{s:pmp}

\begin{figure}
    \centering
    \hspace{-0mm}\includegraphics[width=1.\textwidth]{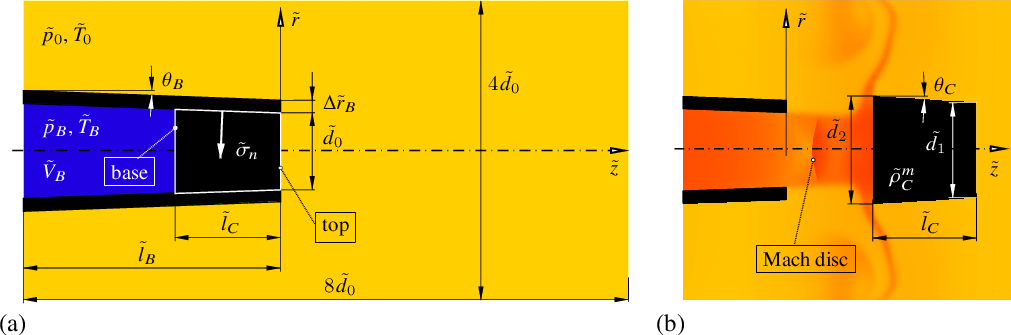}
	\caption{
        (a) Computational domain of coupled system bottleneck--stopper--gas (initial conditions); 
        (b)~detail of density plot at $\tilde{t}=0.5\,$ms; for colour levels of density see figure \ref{Kontur}.}
	\label{Geo}
\end{figure}

For what follows, we tacitly refer to the sketch in figure \ref{Geo} throughout. The physical system considered consists of four parts, subject to a reasonable and self-consistent level of modelling abstraction, in good agreement with real-life observations \citep{Lietal19} and outlined in the following.

\subsection{Basic assumptions}\label{ss:ba}

The physical system considered consists of four subsystems, representing the different materials involved. Its first part represents its boundary, formed by the perfectly rigid, impervious, axisymmetric, relatively slender bottleneck, typically made of glass. The accordingly impervious axisymmetric stopper made of cork forms the second. The third is given by the expanding gas, a solution of CO$_2$, slowly degassed by the liquid champagne and so accounting for its pressurisation prior to the opening of the bottle, air, alcohol and water vapour, which propels the stopper out of the bottle in the axial direction. It is confined by the surface of the contained champagne, during the process considered taken as a stationary, plane and impermeable boundary, and before the opening sealed by the stopper. For what follows, the bottleneck is then precisely defined as the void portion of the closed bottle between its opening and the liquid level. The initially quiescent ambient air represents the fourth subsystem. This and the expanding gas are taken as Newtonian ideal gases having constant specific heats, heat conductivities and dynamic viscosities (perfect gas), all of negligibly different magnitudes. Correspondingly, they form non-distinguishable phases the mixing of which is entirely ignored as are the temperature dependences of those quantities. Even though reasonable for the gas phase, this simplification admittedly disregards the phase transitions observed by \cite{Lietal19}.

The actual opening of the bottle effectively means the rapid overpowering of stiction (self-retention) between the stopper and the bottleneck, compressing the former radially, in favour of much lower sliding friction setting in. The so triggered force imbalance has the gas expand and starts the proper motion of the stopper. Here this very first stage of the opening process is abstracted as instantaneous with respect to the resolved time scales. It thus provides the initial condition. Hence, the stopper is initially accelerated by the gas but decelerated by wall friction. The expanding gas remains contained in the bottle as long as the stopper has not entirely left the bottle. Simultaneously, it starts to move the outer air. Once it has left the orifice behind, it rapidly expands towards its original (decompressed) shape as it is pushed further by the now forming (under-expanded) gas jet and the thereby generated air flow. The gas and the air completely dissolve into one another, and the resulting flow and the motion of the stopper are presumed to remain axisymmetric. 

We neglect the characteristic `dome' of a stopper for champagne bottles, which in combination with the frustum at its base give it the characteristic mushroom-like appearance. However, this renders its discretisation unnecessarily complex whilst taking into account that the drag-increasing bumper would not alter substantially the fascinating flow patterns between the stopper and the bottle opening, tied in with the emergence of the Mach disc and fading out once the stopper is sufficiently remote from the opening. That surprisingly complex shock regime lies at the heart of our numerical analysis.

Before we specify the modelling of the four subsystems and their interactions, we summarise the essential assumptions and simplifications (i)--(vi):
\begin{enumerate}[(v)]
    \item[(i)] The ambient air and the pressurised gas mixture are indistinguishable single-phase ideal gases, characterised by constant (temperature- and pressure-independent) thermophysical properties (hence, the pressurised gas treated as air).
    \item[(ii)] The bottleneck is beheld as a slender truncated cone, tapered towards the bottle opening, with sufficient accuracy.  The mushroom-shaped cap of the stopper and the fastening wire cage (muselage) are both disregarded (as being irrelevant for the process under consideration).
    \item[(iii)] The expanded (unloaded) stopper is also taken as a slender truncated cone, even though real sparkling wine corks are so-called agglomerated ones, composed of up to three rings of slightly different radii \citep{Ma12}. 
    \item[(iv)] The cork used is typically modelled as a homogeneous and isotropic hyperelastic material of negligibly small lateral contraction: for this special case of an Ogden material model see \cite{Saetal22}; Fernandes, Pascoal \& Alves de Sousa (2014), and references therein; \citealp{Og97,Di06}. Here both the weak visco-elasticity usually found to be at play and the degradation of elasticity due to long-time ageing are ignored. Since the compressive reactive force the weakly tapered bottleneck exerts onto the stopper sliding along it appears to be much larger than all the other forces at play, the stopper can deform only radially while it remains rigid in the axial direction.
    Therefore, it assumes its original undeformed shape once its reversible decompression has stopped after a finite, properly defined relaxation time.
    \item[(v)] We \textit{model} the instantaneous expansion of the stopper by taking it as cylindrical when it has just left the bottle and on the basis of the estimated characteristic time scale at play rather than founding it more rigorously. This seems admissible given the relatively short duration of the expansion process.
    \item[(vi)] We neglect any (unavoidable) leakage of gas out of the bottle as long as the stopper has not completely been released from it. That is, we may assume dry sliding (Coulomb) friction between the cork and the glass of the bottle.
\end{enumerate}

Furthermore, the following conventions prove sensible. Tildes indicate dimensional quantities while otherwise their non-dimensional form is used; the subscript $0$ indicates the reference state of the quiescent ambient air; the subscripts $B$ and $C$ indicate the properties of respectively the bottleneck, including the pressurised gas initially at rest and provided by \citealp{Lietal19}, and the cork stopper.

\subsection{Geometry of bottleneck, unloaded stopper and computational domain}\label{ss:gbu}

We introduce cylindrical coordinates $\tilde{r}$ and $\tilde{z}$, radially from and along the axis of the bottle outwards from its opening, respectively. Let us first describe the two truncated cones representing the bottleneck and the fully relaxed stopper as inferred from figure \ref{Geo}. The bottleneck of an axial extent $\tilde{l}_B$ is slightly tapered under an inclination angle $\theta_B$ towards its opening having a diameter $\tilde{d}_0$; this defines its void volume $\tilde{V}_B$. The accordingly shaped relaxed stopper of an axial length $\tilde{l}_C$ has an angle of taper $\theta_C$ and the diameters $\tilde{d}_2$ at its base (rear) end and $\tilde{d}_1\,(<\tilde{d}_2)$ at its top (front) end, this initially exposed to the ambient air. The (average) thickness $\Delta\tilde{r}_B$ of the bottle glass completes the geometry of the systems bottleneck and stopper. 

Hereafter, all lengths are advantageously made non-dimensional with $\tilde{d}_0$. At first, $r=\tilde{r}/\tilde{d}_0$, $z=\tilde{z}/\tilde{d}_0$, $d_{1,2}=\tilde{d}_{1,2}/\tilde{d}_0$, $l_B=\tilde{l}_B/\tilde{d}_0$. Using $V_B=\tilde{V}_B/\tilde{d}_0^3$, we then introduce the (small) slopes
\begin{equation}
  a_B=\tan{\theta_B}=\frac{1}{l_B} \biggl(\sqrt{\frac{3\,V_B}{\pi\, l_B}-\frac{3}{16}}-\dfrac{3}{4}\biggr)\,,\quad 
  a_C=\tan{\theta_C}=\frac{d_2-d_1}{2\,l_C}\,.
  \label{abac}
\end{equation}
The first relationship herein is needed to complete table \ref{t:ginp}, presenting the geometrical input data together with a mean value $\tilde{\rho}_C^m$ of the density of fully relaxed and dry cork \citep{Sa20}. We obtained the geometrical data by measuring customary sparkling-wine bottles and the associated stoppers, apart from the pairing for the larger value of $\tilde{V}_B$, which we could not verify. However, only this and the value of $\tilde{d}_0$ were proposed by \cite{Lietal19}. We extracted the value of $\theta_B$ from the measured one of $\tilde{l}_B$ for $\tilde{V}_B=20\,$ml and \eqref{abac}. In order to accomplish the task of comparing our simulations with the experiments, we then stipulated a congruent bottleneck and thus the same value of $a_B$ for $\tilde{V}_B=25\,$ml. The different liquid level $\tilde{l}_B$ then ensues as the single real root of the arising cubic satisfied by $l_B$ in \eqref{abac}. We also note the so found inclination angles $\theta_B\doteq 2.24$° and $\theta_C\doteq 3.43$°.

\begin{table}
 \centering
 \begin{tabular}{cccccccc}
  \toprule
  $\Delta\tilde{r}_B$ (mm) & $\tilde{d}_0$ (cm) & $\tilde{d}_1$ (cm) & $\tilde{d}_2$ (cm) & $\tilde{l}_C$ (cm) & $\tilde{l}_B$ (cm) & $\tilde{V}_B$ (ml) & $\tilde{\rho}_C^m$ (kg/m$^3$) \\[2pt] 
  \midrule
  $3.015$ & $1.8$ & $2.3$ & $2.6$ & $2.5$ & $6.10/7.28$ & $20/25$ & $240$ \\[2pt] 
  \bottomrule
 \end{tabular}
 \bigskip
 \caption{Geometrical input quantities, cf.\ figure \ref{Geo}, and density of relaxed cork (all suitably rounded).}
 \label{t:ginp}
\end{table}

The axisymmetric computational domain is given by $0\leq r\leq 2$ and $0\leq z+l_B\leq 8$. Numerical tests with enlarged domains, at the expense of considerably increased computational costs, suggest the chosen dimensions as being sufficiently large (cf.\ the discussions regarding the spatial resolution in \S\,\ref{s:rpd}).

\subsection{Gas/air flow}\label{ss:gaf}

According to the premise (i) in \S\,\ref{ss:ba}, we introduce the dynamic viscosity $\tilde{\eta}$, the heat conductivity $\tilde{\lambda}$, the mass-specific gas constant $\tilde{R}$ and the heat capacity ratio $\kappa$ for air under standard conditions as the required, uniform gas properties. Let the ambient air be initially at rest under the uniform pressure $\tilde{p}_0$ and the temperature $\tilde{T}_0$. From the common relations holding for an ideal gas, $\tilde{c}_{\!p}=\tilde{R}\kappa/(\kappa-1)$ is its specific heat capacity at constant pressure, $\tilde{\rho}_0=\tilde{p}_0/(\tilde{R}\tilde{T}_0)$ its density at rest, $\tilde{c}_0=(\kappa\tilde{R}\tilde{T}_0)^{1/2}$ the associated isentropic speed of sound and $\tilde{e}_0=\tilde{\rho}_0 \tilde{c}_{\!p}\tilde{T}_0/\kappa=\tilde{p}_0/(\kappa-1)$ the associated density of the internal energy. We then typically have $\tilde{R}\doteq 287.058$\,J/(kg\,K) and $\kappa\doteq 1.4$. These data together with the characteristic, fixed inner diameter $\tilde{d}_0$ of the cross-section of the bottle opening, adopted by \cite{Lietal19}, establish the reference values for all the remaining input quantities. As the data put forward by \cite{Lietal19} and \cite{Beetal22} shall validate our simulation results, we only consider the two combinations of the pressure $\tilde{p}_B$ and the temperature $\tilde{T}_B$ measured by \cite{Lietal19} and identifying the state of contained gas initially at rest. All the essential input data and their values are summarised in table \ref{t:tpinp} and the resulting four combinations addressed by our simulations in table~\ref{t:cases}.
\begin{table}
 \centering
 \begin{tabular}{cccccccc}
  \toprule
  $\tilde{p}_0$ (bar) & $\tilde{T}_0$ ($^\circ$C) & $\tilde{p}_B$ (bar) & $\tilde{T}_B$ ($^\circ$C) & $\tilde{\eta}$ (Pa\,s) & $\tilde{\lambda}$ $\bigl(\mbox{W/(m\,K)}\bigr)$ & $\tilde{c}_{\!p}$ $\bigl(\mbox{J/(kg\,K)}\bigr)$ & $\tilde{c}_0$ (m/s) \\[2pt] 
  \midrule
  $1.013$ & $20$ & $7.5$ / $10.2$ & $20$ / $30$ & $1.827\times 10^{{}-5}$ & $0.0262$ & $1004.70$ & $343.24$ \\[2pt] 
  \bottomrule
 \end{tabular}
 \bigskip
 \caption{Thermophysical reference data (suitably rounded, slashes separate values of $\tilde{p}_B$ and $\tilde{T}_B$).}
 \label{t:tpinp}
\end{table}
\begin{table}
 \centering
 \begin{tabular}{c|cccc}
  \toprule
  case ($\#$): $\tilde{V}_B$ (ml), $\tilde{T}_B$ (°C) & (A): $20$, $20$ & (B): $25$, $20$ & (C): $20$, $30$ & (D): $25$, $30$ \\[2pt]
  \bottomrule
 \end{tabular}
 \bigskip
 \caption{The combinations \emph{(A)--(D)} of $\tilde{V}_B$ and $\tilde{T}_B$ used in the simulations.}
 \label{t:cases}
\end{table}

Consequently, we consider the local values of five flow quantities, depending on space and time: the fluid velocity $\mathbf{v}$, the fluid density $\rho$, fluid pressure $p$, fluid temperature $T$ and the density of its total energy (the sum of internal and kinetic energy),
\begin{equation}
 e_t=p+\kappa(\kappa-1)\rho|\mathbf{v}|^2\!/2\,,
 \label{et}
\end{equation}
already made non-dimensional with respectively $\tilde{c}_0$, $\tilde{\rho}_0$, $\tilde{p}_0$, $\tilde{T}_0$ and $\tilde{e}_0$. Accordingly, the time $t$ is in natural manner made non-dimensional with the basic time scale $\tilde{d}_0/\tilde{c}_0\doteq 0.0524\,\rm{ms}$. As seen from the particular dimensionless form of $e_t$ in \eqref{et}, $p$ also ensues as the density of the internal energy. Those flow quantities satisfy the thermal equation of state for an ideal gas,
\begin{equation}
 p=\rho T\,,
 \label{tes}
\end{equation}
and the resulting full set of the non-dimensional Navier--Stokes equations. These describe conservation of mass, \eqref{ce}, of momentum, \eqref{me}, and of the total energy density $e_t$, \eqref{ee}, all conveniently written in coordinate-free divergence form:
\begin{eqnarray}
 & \partial_t\rho+\nabla\!\bcdot\!(\rho\mathbf{v})=0\,, &
 \label{ce}\\[2pt]
 & \partial_t(\rho\mathbf{v})+\nabla\!\bcdot\!(\rho\mathbf{v}\mathbf{v})= 
 -\kappa^{-1}\nabla p+\Fr^{-2}\rho\,\mathbf{e}_g+
 \Rey^{-1}\nabla\!\bcdot\!\mathbf{\Sigma}\,, &
 \label{me}\\[1pt]
 & \partial_t e_t+\nabla\!\bcdot\!(\mathbf{v}e_t)=\kappa(\kappa\!-\!1)\Bigl(
 -\kappa^{-1}\nabla\!\bcdot\!(\mathbf{v}p)+\Fr^{-2}\rho\,\mathbf{e}_g\!\bcdot\!\mathbf{v}+
 \Rey^{-1}\nabla\!\bcdot\!\bigl[\mathbf{\Sigma}\!\bcdot\!\mathbf{v}+(\kappa-1)^{-1}\Pra^{-1}\nabla T\bigr]\Bigr)\,. &
 \label{ee}
\end{eqnarray}
Herein, $\mathbf{\Sigma}$ denotes the tensor of the viscous Cauchy stresses for a Newtonian fluid and
\begin{equation}
 \Fr=\tilde{c}_0\big/\sqrt{\tilde{g}\tilde{d}_0}\doteq 817\,,\quad 
 \Rey=\tilde{c}_0\tilde{d}_0\tilde{\rho}_0/\tilde{\eta}\doteq 4.07\times 10^5\,,\quad 
 \Pra=\tilde{\eta}\tilde{c}_{\!p}/\tilde{\lambda}\doteq 0.701
 \label{ndg}
\end{equation}
respectively the Froude, Reynolds and Prandtl numbers as the key groups at play. In the first, $\tilde{g}$ denotes the (constant) scalar gravitational acceleration, $\tilde{g}\doteq 9.81\,\rm{m/s^2}$, acting in the direction of some (constant) unit vector $\mathbf{e}_g$. The figures of these three parameters then follow from the aforementioned input values (see tables~\ref{t:ginp} and \ref{t:tpinp}). The equations \eqref{et}--\eqref{ee} govern the aforementioned five flow quantities in full, apart from the required boundary, coupling and initial conditions discussed below.

Most importantly, $\Fr$ and $\Rey$ appear to be so (predictably) large that we may safely consider the limits $\Fr\to\infty$ and $\Rey\to\infty$ in \eqref{me} and \eqref{ee}. Then \eqref{ce}--\eqref{ee} reduce to the Euler equations governing the adiabatic and inviscid flow of a weightless ideal gas. To specify them for our axisymmetric situation, we introduce the $r$- and $z$-components $u$ and $w$, respectively, of $\mathbf{v}$. Then \eqref{ce}, the two scalar momentum equations for the $r$-and the $z$-direction resulting from \eqref{me} and the energy equation \eqref{ee} give respectively
\begin{align}
 \partial_t\rho+\partial_r(\rho u)+\partial_z(\rho w) &= -\rho u/r\,,
 \label{cee}\\[1pt]
 \partial_t(\rho u)+\partial_r(\rho u^2+p/\kappa)+\partial_z(\rho u w) &= -\rho u^2/r\,, 
 \label{meer}\\[1pt]
 \partial_t(\rho w)+\partial_z(\rho w^2+p/\kappa)+\partial_r(\rho u w) &= -\rho u w/r\,,
 \label{meez}\\[1pt]
 \partial_t e_t+\partial_r\bigl(u[e_t+(\kappa\!-\!1)p]\bigr)+
 \partial_z\bigl(w([e_t+(\kappa\!-\!1)p]\bigr) &= -u[e_t+(\kappa\!-\!1)p]/r\,.
 \label{eee}
\end{align}
The source terms on the right sides of \eqref{cee}--\eqref{eee} are typically due to continuity in the radial direction. Hence, special care is required for resolving the flow near the axis of symmetry $r=0$. We are interested in finding $q(z,r,t)$ where $q$ stands for any of the dependent variables $\rho$, $\rho u$, $\rho w$, $e_t$ and $p$. Eliminating $p$ with the aid of \eqref{et} closes the system of equations \eqref{cee}--\eqref{eee}. Specifically, $e_t+(\kappa\!-\!1)p$ in  \eqref{eee} is replaced by $\kappa[e_t-(\kappa-1)^2\rho|\mathbf{v}|^2\!/2]$. Governing the four remaining quantities, these equations are then ripe for their proper numerical treatment. Most importantly, their divergence form together with \eqref{et} allows for their weak solutions, i.e.\ the capturing of shock waves. The temperature field $T$, the local speed of sound $c=T^{1/2}$, non-dimensional with $\tilde{c}_0$, and the local Mach number $M=(u^2+w^2)^{1/2}\!/c$ can be calculated \textit{a posteriori}, using \eqref{tes}. It is noted that $\kappa$ is the only material-specific parameter entering the problem.

The inviscid-flow description reduces the kinematic boundary conditions to those of axial symmetry and the
impermeability of the gas/liquid interface and the bottle wall:
\begin{equation}
 u(z,0,t)=w(-l_B,r,t)=0\quad (0<r<1/2)\,,\qquad
 \mathbf{v}\bcdot\mathbf{n}_B=0\quad \text{on the bottle surface}\,,
 \label{bcw}
\end{equation}
where $\mathbf{n}_B$ denotes a unit normal of the bottle wall at its respective position. The typical outflow conditions prescribed on the remaining (outer) boundaries of the computational domain complete the boundary conditions. We remark that the neglect of heat conduction and dissipation and the associated boundary layers renders the temperatures at the solid surfaces as just resulting from convection. Therefore, the (questionable) insulation conditions employed by \cite{Beetal22} are irrelevant in our inviscid approach.

The initial conditions are posed at some $t=t_0$ as we choose $t=0$ advantageously as that instance of time when the stopper has \emph{just entirely passed} the bottle opening. That is, $z<0$ and $t<0$ indicate the regime of the sliding/compressed stopper that still seals the bottle, whereas $z>0$ and $t>0$ that of the freely moving released stopper and expanding jet. Hence, for
\begin{equation}
 t=t_0\;(<0)\colon\;\; u=w=0,\;\;\;
 \begin{cases}  
     \;\rho=p=e_t=1 &\;\; \text{(ambient air)}, \\[2pt]
     \;\rho=\tilde{\rho}_B/\tilde{\rho}_0,\;\; p=e_t=\tilde{p}_B/\tilde{p}_0 &\;\; \text{(pressurised gas)}
 \end{cases}
 \label{ic}
\end{equation}
cf.\ \eqref{et}. The particular value of $t_0$ is extracted from the output data of the calculation initialised by \eqref{ic}.

\subsection{Stopper--bottleneck and gas--stopper interaction}\label{ss:sbi}

We first introduce the radial position $r=R(z,t)$ of the surface of the stopper, the axial one $z=Z(t)$ subject to $Z(0)=0$ of its base and the radial ones of the bottleneck, $r=r_B(z)$, and of the relaxed stopper in our resting frame of reference, $r=r_C(z-Z)$. From \eqref{abac},
\begin{equation}
    r_B(z)=1/2-a_B\,z\,,\quad r_C\bigl(z-Z(t)\bigr)=d_2/2-a_C\bigl(z-Z(t)\bigr)\,.
    \label{rbrc}
\end{equation}

The fluid--stopper interaction is facilitated by virtue of the kinematic coupling condition and the dynamic one given by the equation of axial motion of the stopper:
\begin{gather}
    (\mathbf{v}-\mathbf{v}_C)\bcdot\mathbf{n}_C=0\quad \mbox{on the stopper's surface}\,,
    \label{kincc}\\[1mm]
    \kappa\,m_C\ddot{Z}=(F_b-F_t-F_B-F_{ls})(t)\,,\quad m_C=\tilde{\rho}_C^m\tilde{V}_B/(\tilde{\rho}_0 \tilde{d}_0^3)\,.
    \label{ddZ}
\end{gather}
Here $\mathbf{v}_C$ denotes the velocity on and $\mathbf{n}_C$ a unit normal to the stopper surface at the position considered, 
$m_C$ the non-dimensional mass of the stopper (cf.\ table \ref{t:ginp}), and $F_b$, $F_t$, $F_{ls}$ and $F_B$ are respectively the scalar pressure forces at the base, top and the axial components of the forces the fluid and the bottleneck exert at the respective portions of its lateral surface. We term $F_B$ the bottle force as it comprises the $z$-components of the compressive normal force and the tangential Coulomb friction force at play. Given the tapering of the bottleneck and the stopper, see figure \ref{Geo}, all forces in \eqref{ddZ} are taken as positive first.

The forces $F_b$ and $F_t$ have the straightforward explicit representations
\begin{equation}
    F_b(t)=2\pi\int\nolimits_0^{R(Z(t),t)} p\big(Z(t),r,t\big)\,r\di r\,,\quad
    F_t(t)=2\pi\int\nolimits_0^{R(Z(t)+l_C,t)} p\big(Z(t)+l_C,r,t\big)\,r\di r\,.
    \label{Fbt}
\end{equation}
Let $\tilde{\sigma}_n$ and $\sigma_n=\tilde{\sigma}_n/\tilde{p}_0$ denote the stress acting normally onto the portion of the lateral surface of the stopper yet compressed by the bottleneck, $\mu$ the Coulomb friction coefficient. Hence $\mu\sigma_n$ is the tangentially acting frictional stress and $(\sin{\theta_B}+\mu\cos{\theta_B})\,\sigma_n$ the $z$-component of the net stress exerted by the bottleneck on the lateral surface of the stopper yet inside the bottle (see figure \ref{Geo}a). The presumptions (ii) and (iii) in \S\,\ref{ss:ba} allow for approximating both the bottleneck and the stopper by very long cylinders and the static constitutive law proposed in the premises (iv) by taking $\tilde{\sigma}$ as a monotonically increasing function of the local relative compression, $\varepsilon(z,t)$. With $H(\cdot)$ denoting the Heaviside step function,
 \begin{align}
    F_B(t) &= 2\pi\,H(-t)\int\nolimits^0_{Z(t)} \Bigl(-\dfrac{\di r_B}{\di z}+\mu\Bigr)\,\sigma_n\biggl(1-\dfrac{r_B(z)}{r_C\bigl(z-Z(t)\bigr)}\biggr)\,r_B(z)\di z\,,
    \label{FB}\\[2mm]
    F_{ls}(t) &= 2\pi\,H(t)\int\nolimits^{Z(t)+l_C}_{Z(t)} \dfrac{\partial R(z,t)}{\partial z}\,p\big(z,R(z,t),t\big)R(z,t)\di z\,.
    \label{Fls_eq}
\end{align}
Relaxing the assumption of a constant negative slope $\di r_B/\di z$ of the bottleneck in \eqref{FB} signifies the straightforward advancement towards a more general approach. The movement of the stopper decreases its portion inside the bottle and thus $F_B$, to remain zero for $t\geq 0$. We also anticipate that $F_{ls}=0$ for $t<0$: then inside the bottle, the stopper slides along the solid wall; outside, contributions to $F_{ls}$ vanish as its cylindrical shape when it has just passed the opening, assumed in (v) in \S\,\ref{ss:ba}, gives $\partial R/\partial z=0$. 

Closing the relationship $\sigma_n(\varepsilon)$ in \ref{S-s:cls} completes the statement of the problem. \ref{S-s:arf} presents the evaluations of \eqref{FB} and \eqref{Fls_eq} with the localised variation of $R$ examined in \ref{S-s:mse}.

\section{Simulation method and problem-specific innovations of discretisation}\label{s:smp}

The numerical solution of the fluid--structure interaction problem, fully coupling the Euler flow and the dynamics of the stopper as described above, is entirely accomplished within the software environment \textit{Clawpack} (cf.\ \S\,\ref{s:mi}). As a problem-specific numerical challenge, however, the three physical key features, namely the jet flow, the gas--stopper interaction and the sliding between the stopper and the bottle, involve quite disparate time and length scales and force magnitudes. These become critical given the relatively high speeds the stopper attains. Although its surface is found to be slower than the adjacent flow throughout, its considerably fast movement in its normal direction, especially during its radial expansion, raises the vital demand for a tailored calling scheme to achieve a most stable and, therefore, implicit resolution of the gas--stopper coupling. This section addresses the specific care taken in the discretisation of the problem during the intermediate stages of preprocessing and calling.

\subsection{Numerical scheme: overview} \label{ss:Nso}

Let us first recall briefly the layers of \textit{Clawpack}, where the more methodically interested reader is referred to the online software manual by \cite{CP22}, the outline given by \cite{Wa21} and the according references in \S\,\ref{s:mi}.

Godunov's finite-volume discretisation and Roe's linearisation of a Riemann problem of most general type, posed by some hyperbolic system of transport equations expressed in (physically admissible) conservative form, here specified by the Euler equations \eqref{ce}--\eqref{ee}, lie at the core of \textit{Clawpack}. The stable spatial/temporal resolution of weak solutions, i.e.\ discontinuities/shocks, requires the numerically stable evaluation of the local waves and corresponding speeds. This is achieved by constructing the global solution from the waves resolved in the directions normal to the interfaces of adjacent finite-volume cells (\textit{dimensional splitting}) and a \textit{positively conservative} formulation of the actual Riemann solver \citep{HaLaLe83,Ei88,Eietal91}. In practice, the latter requirement can vary greatly in complexity, where the implemented Harten--Lax-van-Leer--Einfeldt (HLLE) scheme by this group of authors provides the probably most prominent one. In each time step, the dependent variables are updated successively via the so obtained scheme and a single call of a second-order Runge--Kutta method. The latter is potentially required to accomplish the time integration of any source terms in the set of equations. Here, their appearance as the right sides of \eqref{ce}--\eqref{ee} resorts to conservation of mass in the radial direction (and thus merely the use of polar coordinates).

\subsection{Boundary and coupling conditions in stable scheme} \label{s-BC}

In the following, we give a synopsis of the numerical implementation, adopting two formal ingredients. \ref{S-s:tdd} puts forward specific technical details. Further information will be provided upon request.

At first, we introduce the solution vector $\mathbf{Q}=(\rho,\rho u,\rho w,e_t)$ and let subscripts indicate the type or the index of the evaluated cell and superscripts the index of the time step considered. In order to start the simulation, every cell must possess initial values, contained in $\mathbf{Q}^0=\mathbf{Q}$ at $t=t_0$. Hence, $\mathbf{Q}^0$ equals either 
$\mathbf{Q}_{\text{in}}^0=\big(\tilde{\rho}_B/\tilde{\rho}_0,\,0,\,0,\,\tilde{p}_B/\tilde{p}_0\big)$ or $\mathbf{Q}_{\text{out}}^0=\big(1,\,0,\,0,\,1\big)$ where the subscripts `in' and `out' refer to the pressurised gas in the bottle and the ambient one, respectively. The cells occupied by the bottle and stopper are filled with non-numeric values so that they are not updated during the subsequent time steps (black regions in figure \ref{Geo}). Secondly, we take into account the method of dimensional splitting in the subsequent outline. It drastically alleviates the discretisation of the kinematic coupling conditions \eqref{kincc} as they can be treated separately for the radial and the axial direction, i.e.\ as if for independent families of one-dimensional waves.

In \textit{Clawpack}, boundary conditions are realised by including \textit{ghost cells} near the edges of the domain and at the interfaces between solid and gaseous phases (for a graphical depiction see Supplement~\ref{S-ss:GpBC}). The kinematic constraints \eqref{bcw} and \eqref{kincc} are treated as reflective walls, and extrapolating boundary conditions simulate an undisturbed flow out of the spatial domain (outflow conditions). The extrapolation condition assigns the first and last internal value to all ghost cells at the left and right or base and top edges, respectively: $\mathbf{Q}_j=\mathbf{Q}_{N_G+1}$, $\mathbf{Q}_{N_l+j}=\mathbf{Q}_{N_l}$ ($j=1,\ldots,N_G$). Here $N_G$ denotes the number of ghost cells and $N_l$ the last internal cell index. With regard to \eqref{kincc}, ghost cells inside the stopper result from mirroring fluid cells at its surface. Velocities are assigned to the ghost cells such that the average value of their and the original flow components perpendicular to the surface match that of its local normal speed ($\Dot{Z}$ or $\Dot{R}$). All other quantities are simply mirrored along the interface. Its index is set to $m-1/2$ consistently where $m$ indicates the cell centre: $\rho_{m-1+j}=\rho_{m-j}$, $p_{m-1+j}=p_{m-j}$,
$\mathbf{v}_{m-1+j}=2\mathbf{v}_{C,m-1/2}-\mathbf{v}_{m-j}$ ($j=1,\ldots,N_G$). While this allocation describes the filling of ghost cells right of a reflective wall, these equations can simply be rearranged to apply to a wall to the left. As a particular novelty, $p$ is mirrored instead of $e_t$, which as part of the solution is the intuitive and traditionally used choice (see the aforementioned literature on \textit{Clawpack}). However, if the method is performed in this classical manner, unacceptable negative values of the pressure inevitably arise inside the HLLE algorithm when the value of the kinetic contribution in~\eqref{et} exceeds a certain threshold.

The maximum value of the variable time step must be smaller than that given by the \textit{CFL} condition \citep{CFL28}: $u\Delta t/\Delta r+w\Delta t/\Delta z\leq C_{\max}$. For explicit time stepping methods $C_{\max} =1$. A grid with quadratic cells is chosen ($\Delta r=\Delta z$) to minimise distortions. A variety of different cell distances and time steps is used in order to assess the consistency of the discretisation and the temporal convergence of the scheme~(\S\,\ref{s:rpd}).

The (continuous) motion of the stopper is resolved for discrete but variable time steps $\Delta t_n=t_{n+1}-t_n$ using Taylor approximations. Setting $\tau=\Delta t_n/\Delta t_{n-1}$, we approximate the base position of the stopper $Z(t_n)$ and the corresponding velocity $\Dot{Z}(t_n)$ with $Z^n$, $\Dot{Z}^n$ as
\begin{align}
    Z^{n+1} &= -\tau\,Z^{n-1}+(1+\tau)\,(Z^n+\ddot{Z}^n\,\Delta t_n\,\Delta t_{n-1}/2)\,, \label{Zn}\\
    \Dot{Z}^{n+1} &= \tau^2\,\Dot{Z}^{n-1}+(1-\tau^2)\,\Dot{Z}^n+\Delta t_n (1+\tau)\,\ddot{Z}^n\,.
    \label{Zdn}
\end{align}

\section{Results, post-processing and discussion}\label{s:rpd}

We now scrutinise the simulated results for the four cases shown in table \ref{t:cases}. The interaction of the stopper with its surrounding fluid heavily influences the movement of the stopper and also the behaviour of the Mach disc. 
The latter is obviously more visible in the cases of the higher initial temperature/pressure, which promotes a greater richness of the flow structure. Since here case (C) is substantiated by our own measurements, we discuss the dynamics of the stopper and some aspects of its measured counterpart \citep{Lietal19} for this reference case in greater detail before we address the comparison of the simulated flows for all situations.

The following enumeration summarises the key flow features in chronological order, where the specific time stamp of these events depend strongly on the considered scenarios (table \ref{t:cases}). As will become evident later, it is sensible to distinguish (roughly) between six steps, referenced by (0) and (a)--(e):
\begin{enumerate}
    \item[(0)] The stopper, initially completely inside the bottle, starts to move out of the opening as a cylinder (see \ref{S-s:mse}).
    \item[(a)] The radial expansion of the stopper starts exactly when it has fully escaped the bottle. It takes approximately $40\,\mu$s throughout to reach its final, decompressed state (\ref{S-s:mse}). The bottle emits the gas through the thin gap forming between the stopper and the bottle as a supersonic, first radially expanding jet. A pressure wave is emitted. 
    \item[(b)] At about the opening radius, a shock emerges at the edge of the jet and wanders radially inwards.
    \item[(c)] The shock finally forms a Mach disc of initially convex shape. The jet, now moving axially with the stopper, drives it away from the opening.
    \item[(d)] Having reached some maximum distance, the Mach disc retracts until it dissolves into a compression wave propagating inside the then diverging bottleneck, with about $M\approx 0.5$, against the gas still out-flowing, also with $M\approx 0.5$. Correspondingly, transonic conditions prevail at the opening. The jet bends and starts to overtake the stopper in the positive axial direction. Also, a second Mach disc can emerge and often exhibits a behaviour similar to the first one. Both Mach discs can experience all phases or simply vanish after a certain period of time.
    \item[(e)] During the (with respect to the simulated fraction of the opening process) end stage, the compression wave may be reflected on the liquid interface and might again steepen to form a shock (Mach disc) propagating outwards. 
\end{enumerate}

We stress that the axisymmetry of the flow cannot be reduced further, not even locally: unlike for the one-dimensional classical Riemann problem referring to a cylindrical shock tube \citep[cf.][]{Sod78}, here the just released stopper inevitably invokes both axial and radial flow variations near the bottle opening. On the other hand, \textit{Clawpack} adopts a quasi-one-dimensional approach by combining the local solutions of the Riemann problem for each dimension (dimensional splitting, cf.\ section \ref{ss:Nso}), and the stopper-free benchmark simulation carried out by \cite{Wa21} is indeed closely related to the shock tube problem.

The grid resolution ranges from $200\times 50$ to $3200\times 800$ cells and the maximum real-time steps from $500\,$ns down to $50\,$ns. Unless stated otherwise, the results are obtained with the latter (finest) spatial and temporal resolutions. According to the finite-volume discretisation, summation of the cell pressures yield their forces and a linear interpolation between the data points the two-dimensional graphs. The contour plots do not employ any interpolation at all. We conveniently provide the reader with dimensional results according to the scalings introduced in \S\,\ref{ss:gaf}: further notable conversions are $\tilde{F}=\tilde{p}_0\tilde{d}_0^2 F$, $\tilde{t}=(\tilde{d}_0/\tilde{c}_0)t$ (subscripts omitted), $\tilde{Z}=\tilde{d}_0\,Z$, $\tilde{\Dot{Z}}=\tilde c_0\,\Dot{Z}$, $\tilde{\ddot{Z}}=(\tilde{c}_0^2/\tilde{d}_0)\,\ddot{Z}$. All data are rounded appropriately.
The Supplementary Material contains two animated videos displaying the simulation of the case above.

\subsection{Interaction of gas with stopper passing bottle opening}\label{ss:igs}

\begin{figure}
	\centering
	\includegraphics[width=0.49\textwidth]{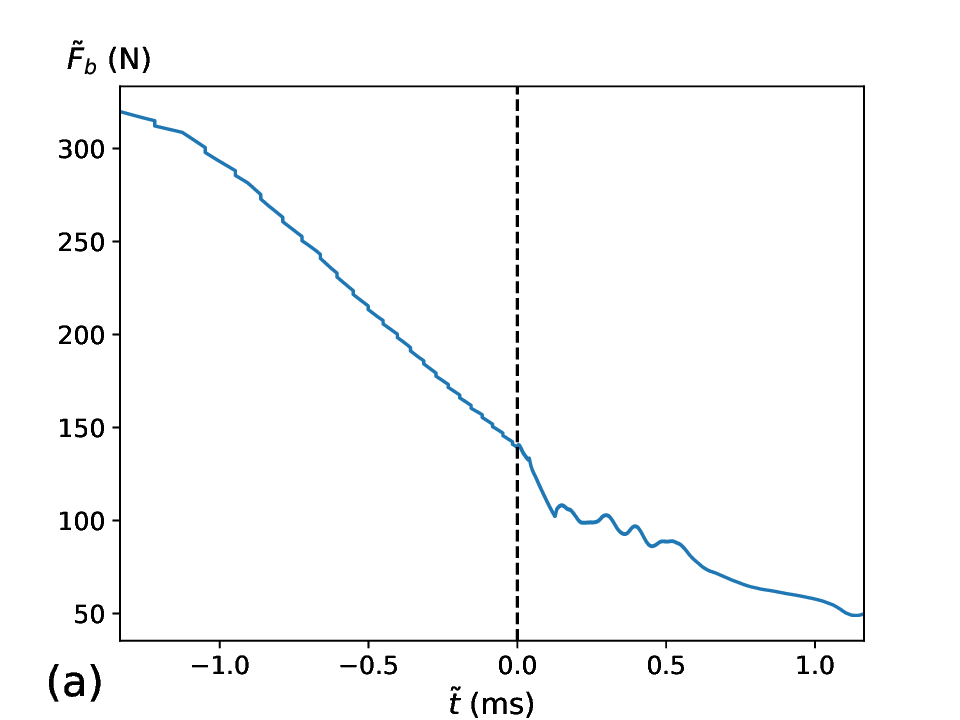}
	\includegraphics[width=0.49\textwidth]{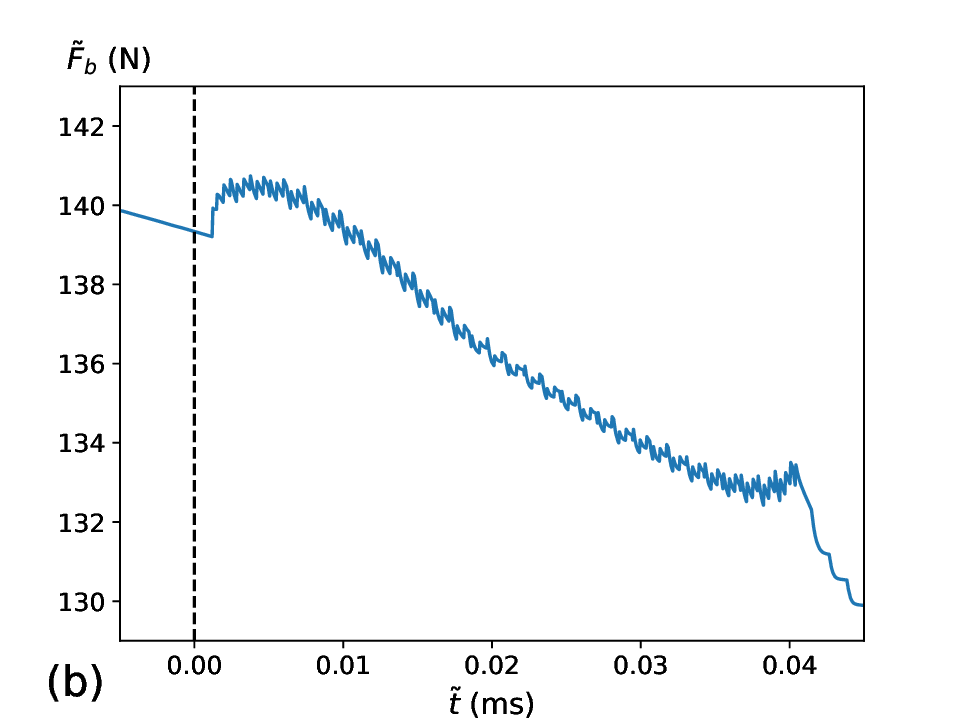}
	\caption{Stopper dynamics: 
        (a)~force $\tilde{F}_b$ at base surface vs.\ $\tilde{t}$; 
        (b)~magnification around $\tilde{t}=0\,$ms.}
        \label{Fl}		
\end{figure}

From the start of the simulation until the complete escape of the stopper out of the bottle, $\tilde{F}_b$ decreases (rather linearly) with time (figure \ref{Fl}a). This fact is the result of two phenomena: the expansion of the gas behind the stopper decreases the density and thus the pressure of the gas; also, the bottleneck and therefore the sliding stopper become narrower as it moves towards the opening, having the base diameter decreased. Its numerical value changes at discrete times. Their distances increase and, in turn, the temporal resolution of $\tilde{F}_b$ becomes more visible the coarser the grid is and the smaller the slope of the bottleneck. During the expansion of the fully released stopper, its base surface rapidly increases, explaining the initial increase of $\tilde{F}_b$ (figure \ref{Fl}b). The following variation of $\tilde{F}_b$ originates in the adopted constitutive behaviour of cork controlling its radial expansion. This is so rapid at the edge of the base that the numerical method seems to partially resolve a spurious flow field generated by the just produced ghost cell: the feedback of the fast flow in the just forming gap separating the bottle and the stopper starts to compensate this, but the newly arising ghost cell triggers its reemergence and, in turn, oscillations. (More generally, the lagging adaption of the flow to a moving surface heralds a potential localised failure of the ghost-cell method once the speeds of its generation/annihilation and of the neighbouring flow become comparable; cf.\ the comments at the begin of \S\,\ref{s:smp}.) 
While the numerical errors/algorithmic instabilities induce small short-scale oscillations, the slower but larger self-sustained waves seen in figure \ref{Fl}a ($0.1$--$0.5\,$ms) have a physical origin. Albeit interfering with the supersonic jet and reflected on the base of the stopper, they are too weak to be resolved in any of the upcoming flow visualisations.

\begin{figure}
	\centering
	\includegraphics[width=0.49\textwidth]{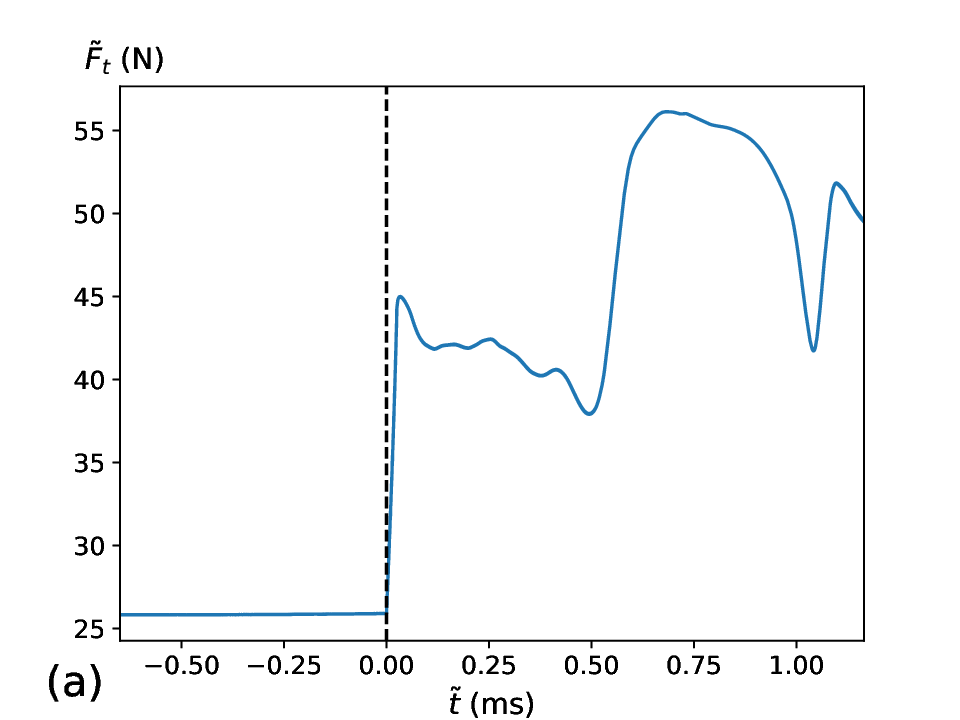}
	\includegraphics[width=0.49\textwidth]{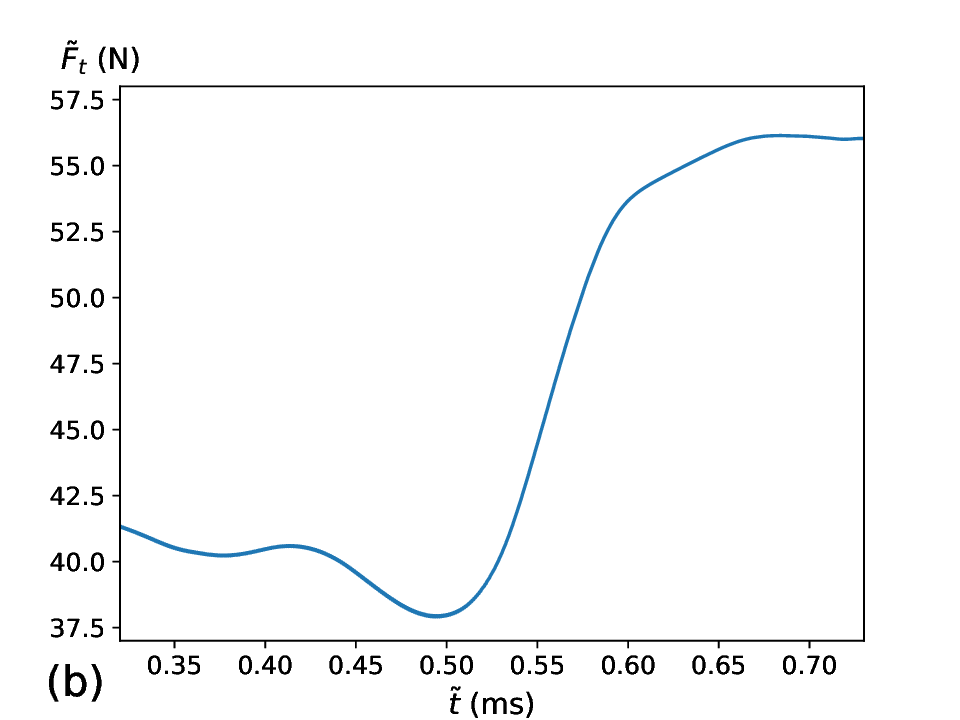}
	\caption{Stopper dynamics: 
        (a)~force $\tilde{F}_t$ at top surface vs.\ $\tilde{t}$;
        (b)~magnification around $\tilde{t}=0.5\,$ms.}
	\label{Fr}	
\end{figure}

\begin{figure}
	\centering
	\includegraphics[width=0.49\textwidth]{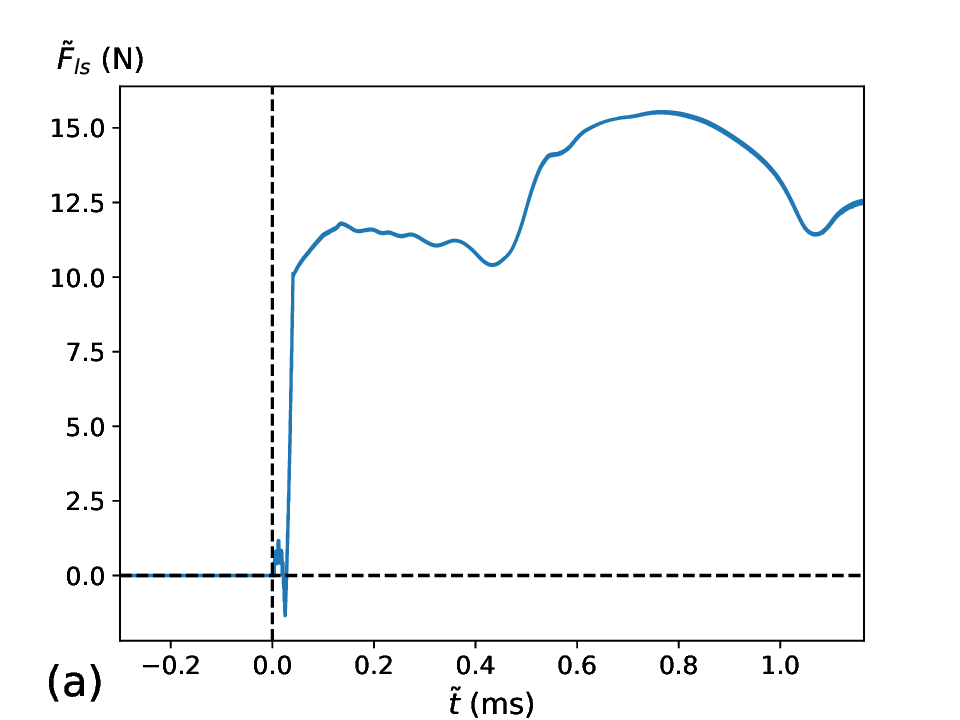}
	\includegraphics[width=0.49\textwidth]{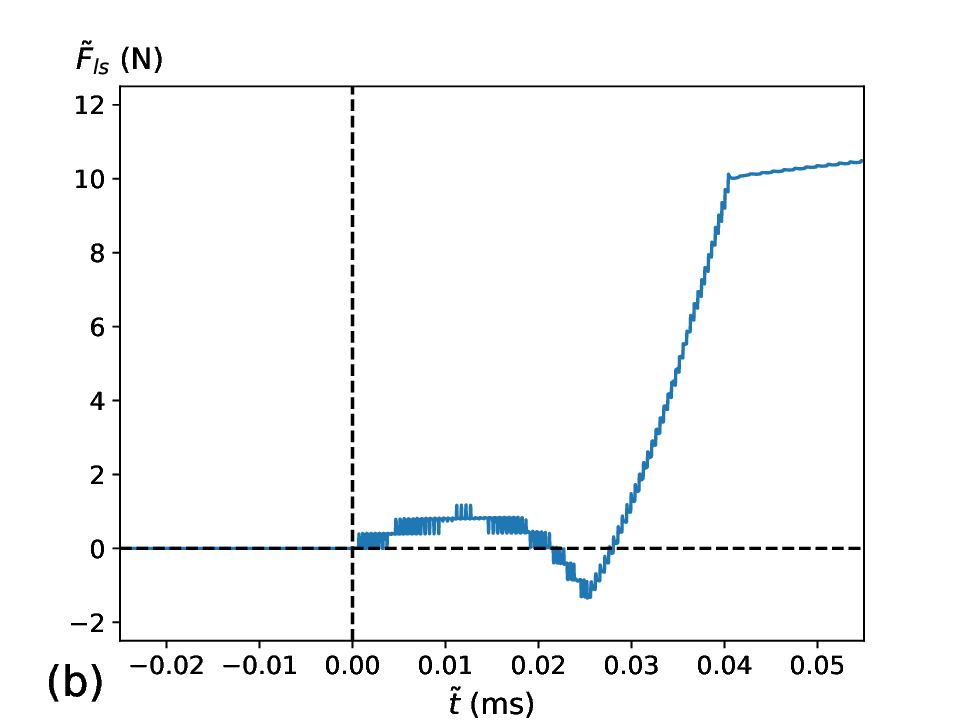}
	\caption{Stopper dynamics: 
        (a)~force $\tilde{F}_{ls}$ at lateral surface vs.\ $\tilde{t}$; 
        (b)~magnification around $\tilde{t}=0\,$ms.}
        \label{Fls}		
\end{figure}

Figure~\ref{Fr}b displays the pressure force at the top surface, $\tilde{F}_t$. Initially, $\tilde{F}_t$ is less than $10\%$ of $\tilde{F}_b$. It remains almost constant as long as the stopper seals the bottle and its top cross section has not expanded. The increase after 0.5\,ms discerned in figure \ref{Fr}b is the result of the leading pressure wave overtaking the stopper (see figure \ref{Kontur}d further below).
Both the pressure and friction forces acting on the lateral surface of the stopper decelerate the object. However, they vary greatly in size and appearance. Here the pressure-induced component $\tilde{F}_{ls}$ is the smallest of the forces resulting from the gas--stopper interaction (figure \ref{Fls}). The base surface area expands quicker than the top one (see \ref{S-s:mse}, figure \ref{vc,R}b), initially resulting in a negative mean slope of the lateral surface and thus reducing the acceleration of the stopper. However, this effect changes shortly after. At about $25\,\mu$s, the top surface has already reached its relaxed state while the base is still expanding, and $\tilde{F}_{ls}$ takes on a negative minimum and stays negative for a short period of time (figure \ref{Fls}b). After the stopper has reached its terminal expanded shape, its axial slope remains equal to $-a_C$ and therefore $\tilde{F}_{ls}$ positive, only oscillating in accordance with the pressure forces acting on the other surfaces.

\begin{figure}
	\centering
	\hspace*{-0.7cm}\includegraphics[width=0.495\textwidth]{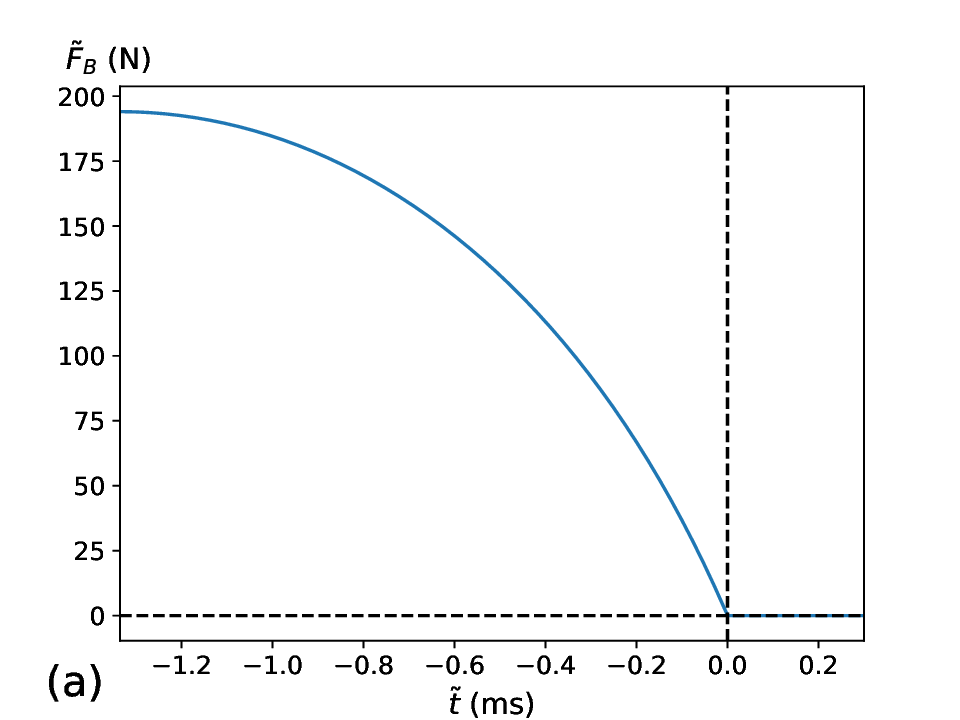}
    \hspace*{0.2cm}\includegraphics[width=0.425\textwidth]{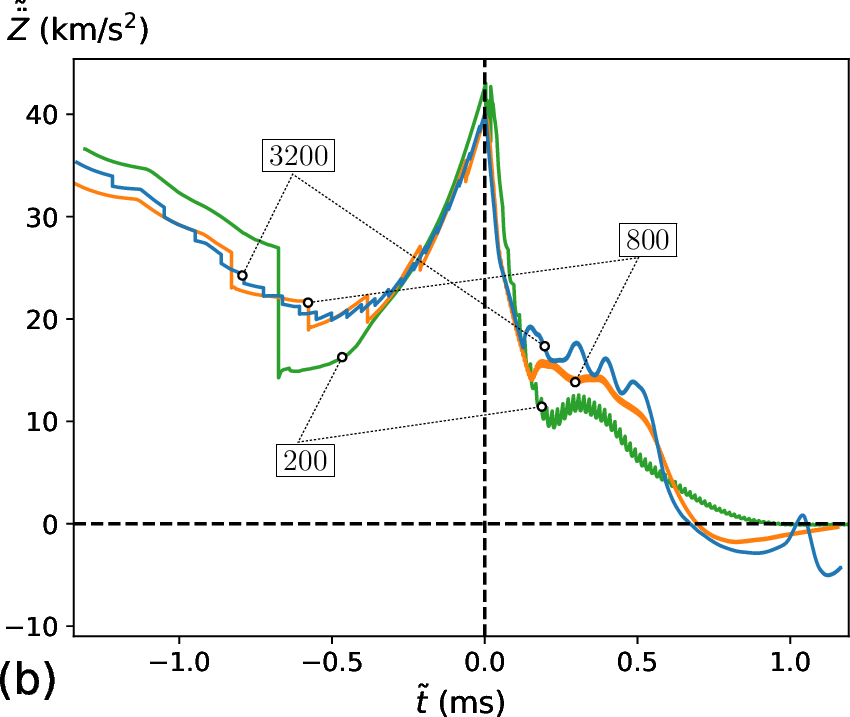}
	\caption{Stopper dynamics: 
        (a)~bottle force $\tilde{F}_B$ and 
        (b)~acceleration $\tilde{\ddot{Z}}$ vs.\ $\tilde t$ for $3200$ (blue), $800$ (orange) and $200$ (green) cells in axial direction.}
        \label{FfZdd}
\end{figure}

The temporal development of the bottle force $\tilde{F}_B$ is plotted in figure \ref{FfZdd}a (see \ref{S-s:arf}). It is key for determining when and even if the stopper will exit the bottle for a given set of initial conditions. Rather surprisingly, the simulations with our originally extrapolated value $0.2$ of $\mu$ and the more sensitive cases where $\tilde{T}_B=20$°C (i.e.\ of a lower initial driving pressure $\tilde{p}_B$) predicted the stopper getting stuck inside the bottleneck. Consequently, we set $\mu$ to $0.15$ in turn so as to grant the realistic scenario of an escaping stopper for all cases considered. The formula \eqref{FBt} is rather lengthy but provides the following straightforward explanation of the initial dynamics of the stopper. Due to the sliding, its lateral surface along the convergent bottleneck steadily decreases, thereby also reducing the value of $\tilde F_B$. Consequently, the bottle force drops to reach zero at $t=0$ and stays at that value for $t>0$. Figure~\ref{FfZdd}b shows the total acceleration calculated from \eqref{ddZ} for three different grid resolutions. Because of their dominant magnitudes compared to those of the other forces at play, only $\tilde F_b$ and $\tilde F_B$ influence the shape of the acceleration significantly. Although both forces decrease over time, $\tilde F_B$ does so more slowly, therefore resulting in a decline of $\tilde{\ddot{Z}}$ until $\tilde{t}\doteq -0.5\,\rm{ms}$, which becomes marked for the highest resolution (blue graph). The minimum value of $\tilde{\ddot{Z}}$ determines if the stopper will get stuck inside the bottle. For this behaviour to occur, the acceleration of the stopper must reach such a negative value that also its speed becomes negative. Due to $\tilde F_B$ remaining zero for all $\tilde t\geq0$ and the gradient of $\tilde F_b$ being negative nearly at every point in time, the maximum value of the acceleration must be at around $\tilde t=0$. After that, the oscillation pattern is the result of all pressure forces combined, which is patently more visible for finer grids when neglecting all the grid oscillations intrinsic to the numerical resolution.

The speed $\tilde{\Dot Z}$ and the base position $\tilde Z$ of the stopper are typically resolved as much smoother than its acceleration $\tilde{\ddot{Z}}$: see figure \ref{v,Z}. While this directly results from the pressure forces that are inherently bound to the discretised spatial domain, the update \eqref{Zdn} of the stopper speed tends to smear out sharp edges. Since $\tilde{\Dot Z}$ increases from case (A) to case (D) (see table \ref{t:cases}, figure \ref{v,Z}), the start time $\tilde t_0$, defining $\tilde Z(0)=0$, increases likewise. A negative acceleration only occurs in the first case, thereby explaining the local minimum of the green graph in figure \ref{v,Z}a. In the case (A), the stopper undergoes an exit speed of around $18\,$m/s, a value closely resembling the experimental data found in \cite{Lietal19} and used in \cite{Beetal22}.
\begin{figure}
	\centering
 \hspace*{-0.5cm}\includegraphics[width=0.43\textwidth]{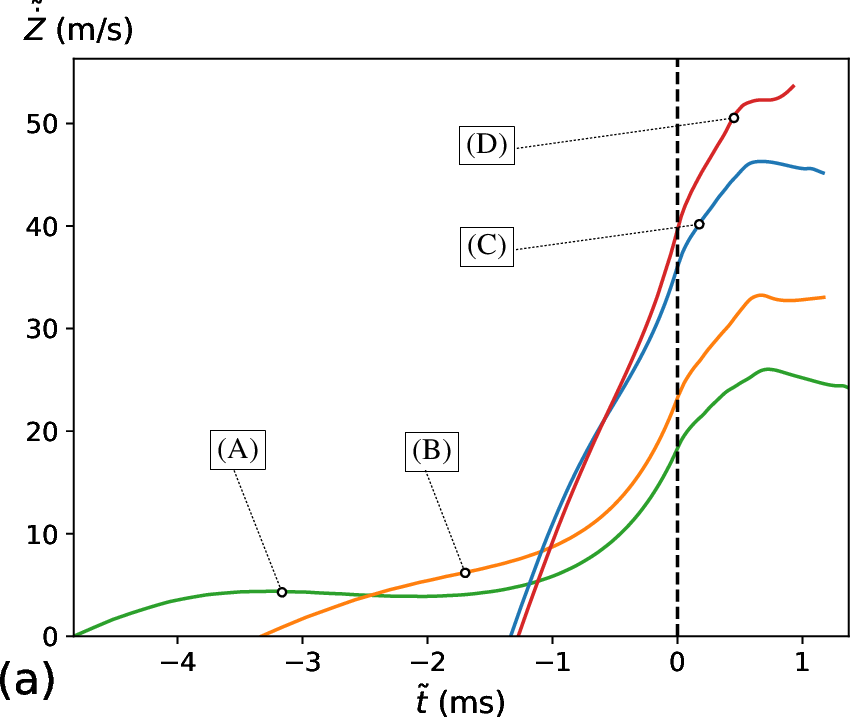}
	\hspace*{0.7cm}\includegraphics[width=0.43\textwidth]{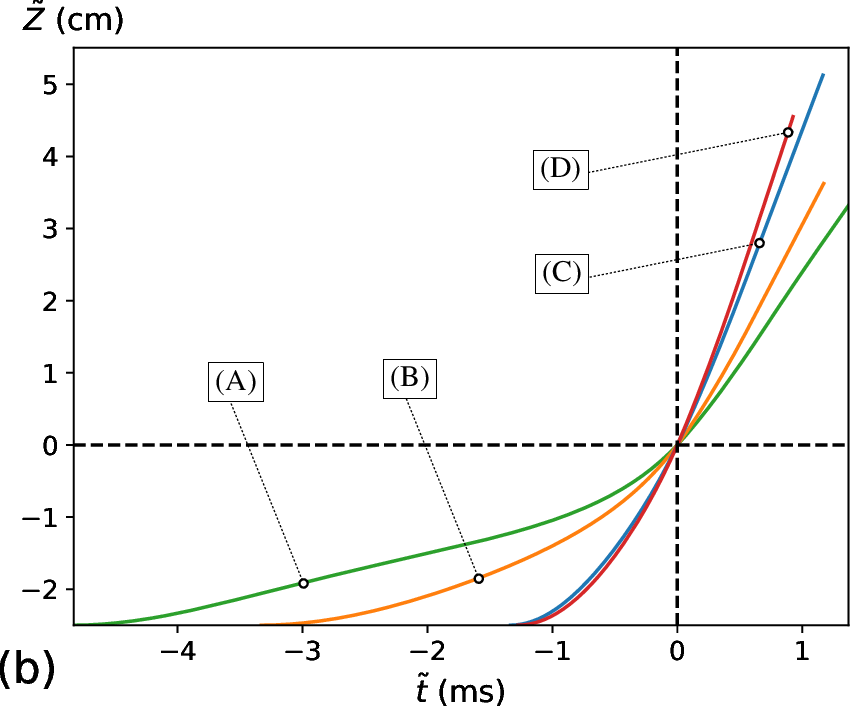}
	\caption{Stopper dynamics: 
        (a)~speed $\tilde{\Dot{Z}}$ and 
        (b)~base position $\tilde{Z}$ vs.\ $\tilde{t}$ for the cases in table \ref{t:cases}: green (A), orange (B), blue (C) and red (D).}
	\label{v,Z}
\end{figure}

\subsection{Resolution and consistency study}

We focus on the long-time dynamics to discuss the independence of our results of the (coupled spatial--temporal) numerical resolution.
As mentioned briefly in section \ref{ss:igs}, the required spatial resolution is determined from figure \ref{FfZdd}b: the physical oscillations must predominate markedly over the numerical (spurious) ones. This is the case for 3200 cells in the axial direction. Any further grid refinement would not be discernable in figure \ref{FfZdd}b but increase the computation time, roughly proportional to the total number of cells in the axial direction squared. As a reference, the simulation was performed on a \textit{Intel\textsuperscript{\,\textregistered}\,Core i9-12900K} processor where case (A) needed the longest computation time of about 24 hrs.

In order to assess the consistency of the simulated data and temporal convergence, we evaluate the accumulated mass of gas that has already exited the bottle in two different ways and compare the results: at first, we introduce $\tilde m_\mathrm{out,I}(\tilde{t})$ as the mass flow through the bottle opening integrated over time from $\tilde t=0$; secondly, we define the same quantity as $\tilde m_\mathrm{out,II}(\tilde{t})=\tilde m_\mathrm{in}(0)-\tilde m_\mathrm{in}(\tilde{t})$, where $\tilde m_\mathrm{in}(\tilde{t})$ is the mass of gas yet contained in the bottle. While the discretisations of the area integral over the mass flux density to obtain $\tilde m_\mathrm{out,I}$ and of the volume integral over $\rho$ to compute $\tilde m_\mathrm{in}$ are analogous to that of the pressure forces, the time integration is based on the Euler method. The so encountered integration error is consistent with that due to the flow simulation. Obviously, $\tilde m_\infty=\tilde m_\mathrm{in}(\infty)$ must equal the mass of gas filling the volume $\tilde{V}_B$ of the bottleneck in the state of full equilibrium (cf.\ figure \ref{Geo}a). Therefore, $\tilde m_\infty$ serves as an appropriate reference quantity for highlighting the approach towards global equilibrium as the non-dimensional time $t$ takes on relatively large values. 

\begin{figure}
	\centering
 \hspace*{-0.1cm}\includegraphics[width=0.475\textwidth]{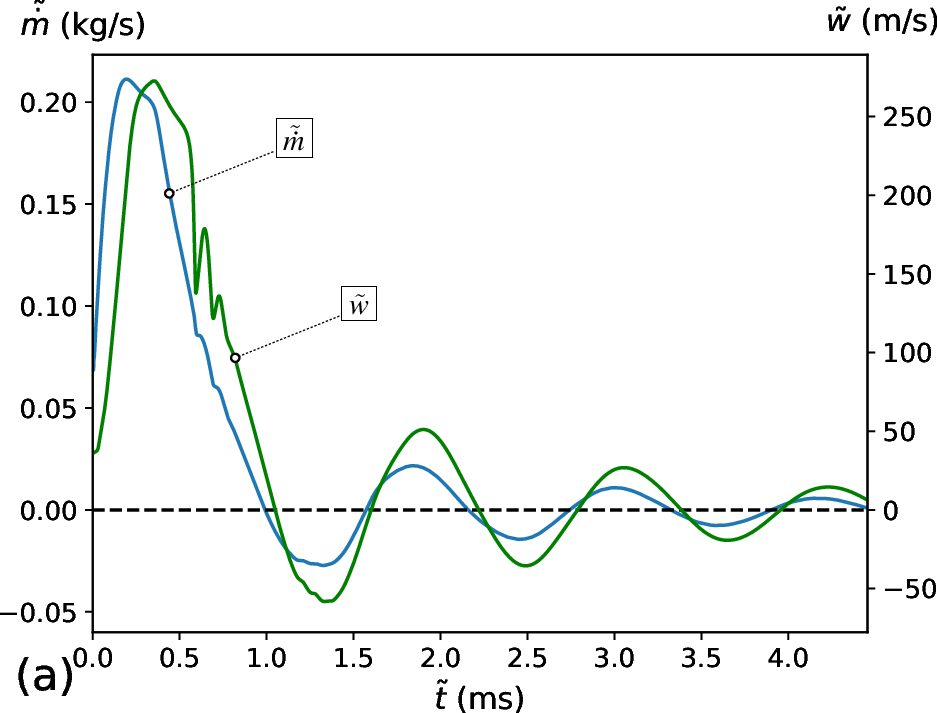}
 \hspace*{0.3cm}\includegraphics[width=0.47\textwidth]{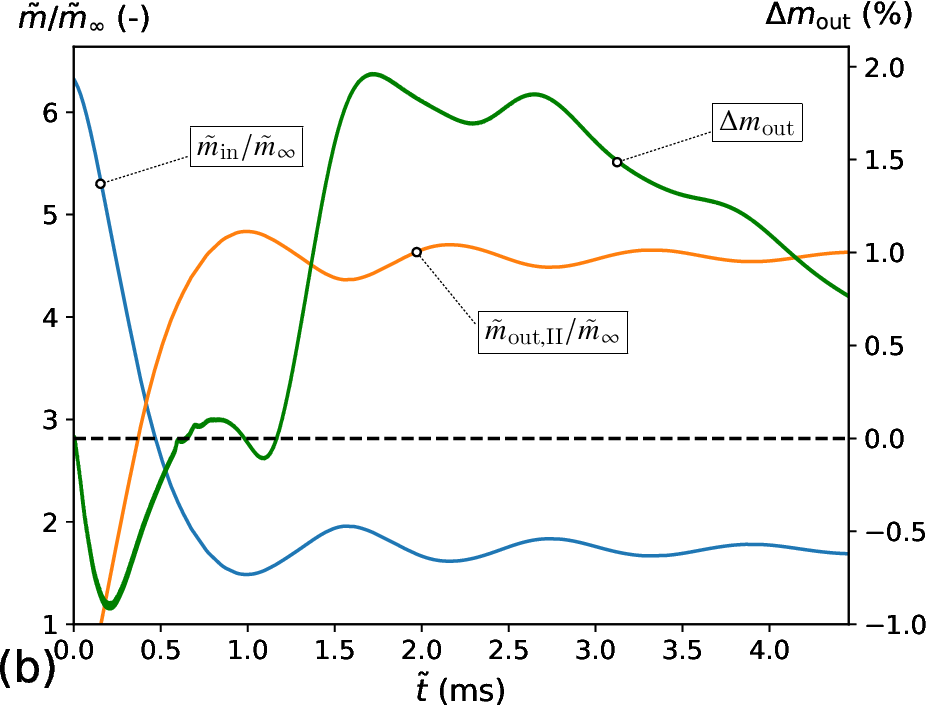}
	\caption{(a) Mass flow $\tilde{\dot m}$ (blue) and axial velocity $\tilde{w}$ at centre of opening (green), 
                 (b) normalised internal $\tilde m_\mathrm{in}/\tilde m_\infty$ (blue) and expelled $\tilde m_\mathrm{out,II}/\tilde m_\infty$ (orange) gas mass and relative error $\Delta m_\mathrm{out}$ (green), all vs.\ $\tilde{t}$ and for $1600\times 200$ cells and 
                 $0\leq\tilde{z}+\tilde{l}_B\leq 16\,\tilde{d}_0$.}
        \label{masses}
\end{figure}

In figure \ref{masses}a, the mass flow through the bottle opening, $\tilde{\dot m}(\tilde t)$, and $\tilde{w}$ at its centre (coordinate origin) are plotted. Figure~\ref{masses}b displays the accumulated masses obtained by the two methods and normalised with $\tilde m_\infty$ as well as their relative difference $\Delta m_\mathrm{out}=[\tilde m_\mathrm{out,I}(\tilde{t})-\tilde m_\mathrm{out,II}(\tilde{t})]/\tilde m_\infty$. This exhibits a maximum value of about $(1.96\pm 0.5)\%\,$, varying only insignificantly with different (satisfactorily high) temporal resolutions. Here the axial extent of the computational domain was doubled but the spatial resolution kept fixed so as to accordingly stretch the time interval in which the movement of the stopper is resolved.

Due to the compressible `sloshing', associated with the occurrence of the Mach discs, all quantities except $\Delta m_\mathrm{out}$ disclose `damped' oscillations. While those seem to be in phase for both $\tilde {\dot{m}}$ and $\tilde w$ in figure \ref{masses}a, the only perturbations in $\tilde w$ occur when the first and second Mach disc enter the bottle. Because the relaxation time is much bigger than the simulation time, $\tilde m_\mathrm{in}/\tilde m_\infty$ in figure \ref{masses}b only deceptively seems to attain a value greater than one for sufficiently large times. However, while the relative error $\Delta m_\mathrm{out}$ exhibits its maximum when the oscillations are most pronounced, it indeed stays below 2\% and dies out for larger values of $\tilde t$, which indicates temporal convergence.

\subsection{Parameter study and comparison with experimental data}\label{ss:psc}

\begin{figure}
	\centering
 \hspace*{-0.3cm}\includegraphics[width=0.465\textwidth]{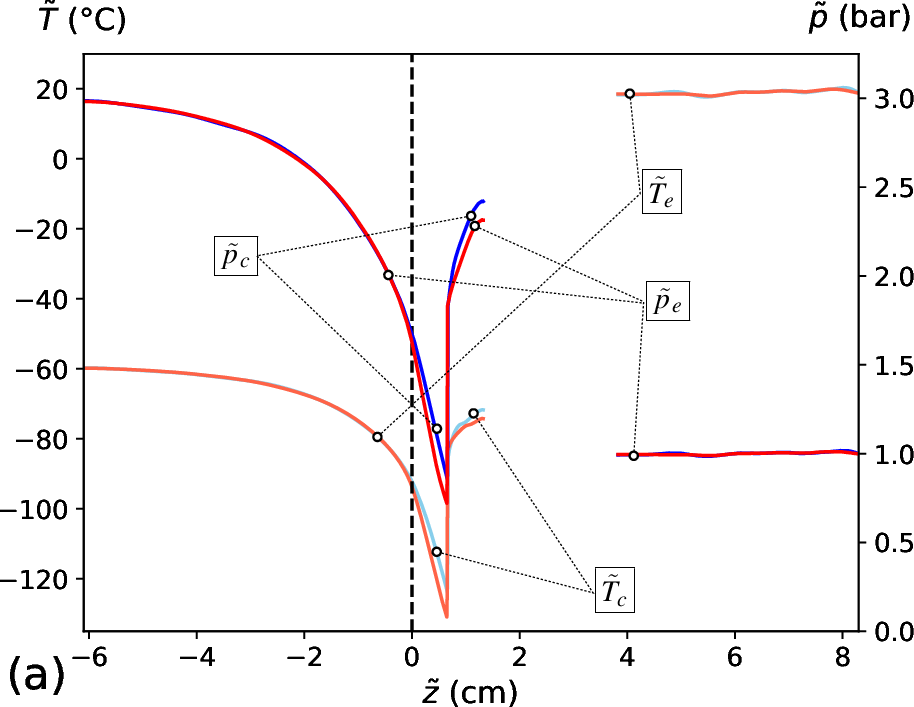}
	\hspace*{0.6cm}\includegraphics[width=0.435\textwidth]{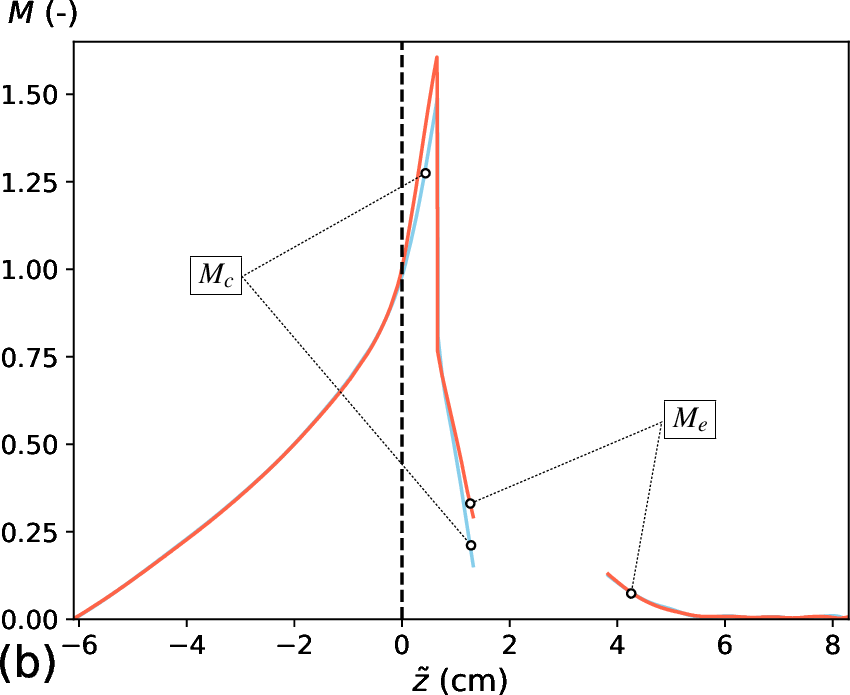}
	\caption{Snapshots of (a) temperature (light colour), pressure (dark colour) and (b) Mach number vs.\ $\tilde z$; both graphs depict quantities at edge (red) and centre (blue) of Mach disc for $\tilde{t}=0.331\,${\rm ms}; the stopper fills the gap between the two branches of the curves.}
        \label{Mach2D}
\end{figure}

A careful detection of any discontinuities in all flow quantities reveals the occurrence of shock waves: see the instantaneous distributions of $\tilde{T}$, $\tilde p$ and $M$ depicted in figure \ref{Mach2D}. While figure \ref{Mach2D}a visualises the isentropic relation between $\tilde{T}$ and $\tilde p$, their minimum values also explain the conditions for the creation of dry ice particles, mentioned by \cite{Lietal19}. Figure~\ref{Mach2D}b on the other hand is used for determining the distances between the Mach discs and the bottle opening (see the supplementary movie ``Movie 1''). Therefore, we analysed $M$ along the $z$-axis and along a parallel line $r=0.25$, i.e.\ centred between this and the outer edge of the bottle opening. It turned out that this value satisfactorily measures the initial edge of the disc. Then the shock positions along these lines, and thereby of the centre and edges of the Mach discs, were determined by two methods. In a first attempt, we considered them where $M$ undergoes its maxima. However, the locations of the (positive) maxima of $-\partial_z M$ (which lie immediately downstream of the maximum values of $M$) allowed for their much more accurate prediction. More precisely, if the value of such a local maximum exceeds a certain threshold, 5.5 in case (A), 5.8 in cases (B)--(D), we assign its $z$-position to that of the Mach disc. Notably, a low threshold allows to capture weaker Mach discs more easily, hence predominantly on coarser grids; but for a certain value, the algorithm falsely detects shock waves where none are physically present.
A further noteworthy feature of the flow extracted from figure~\ref{Mach2D}b are the near-critical conditions around the bottle exit, typically raised by the only slightly converging cross section of the bottleneck.
%
%

\begin{figure}
	\centering
	\includegraphics[width=0.38\textwidth]{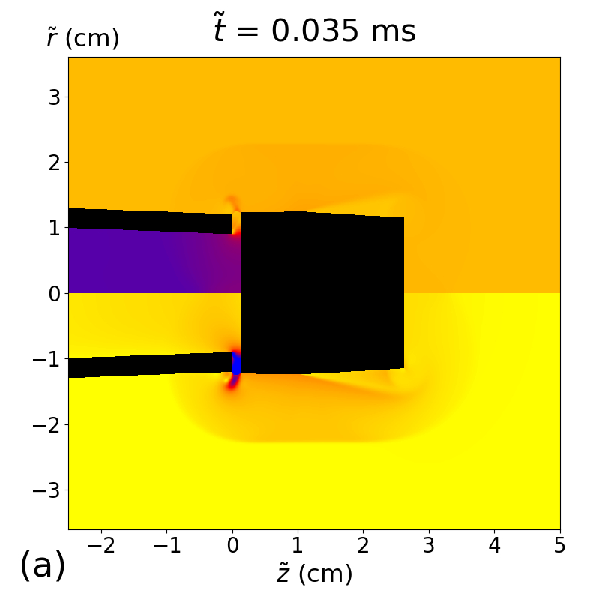}
	\hspace*{1.5mm}
	\vspace*{2mm}
	\includegraphics[width=0.503\textwidth]{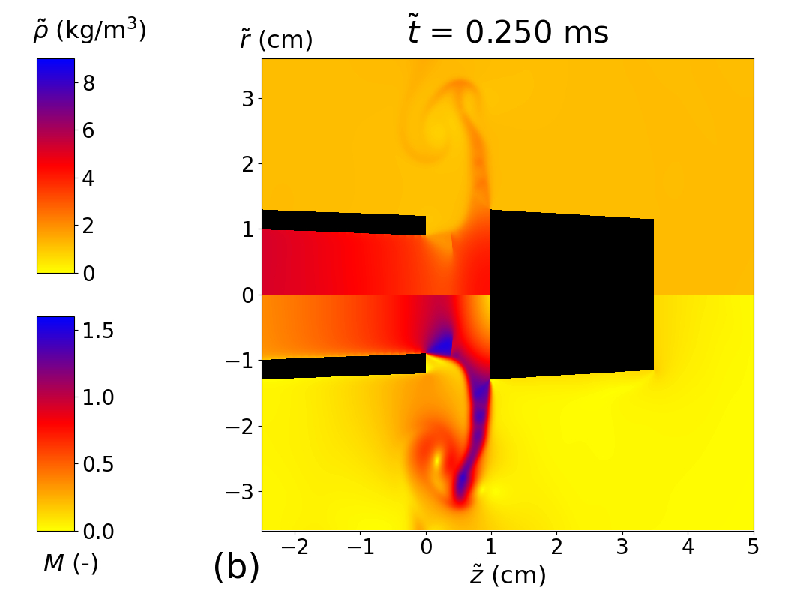}
	\includegraphics[width=0.38\textwidth]{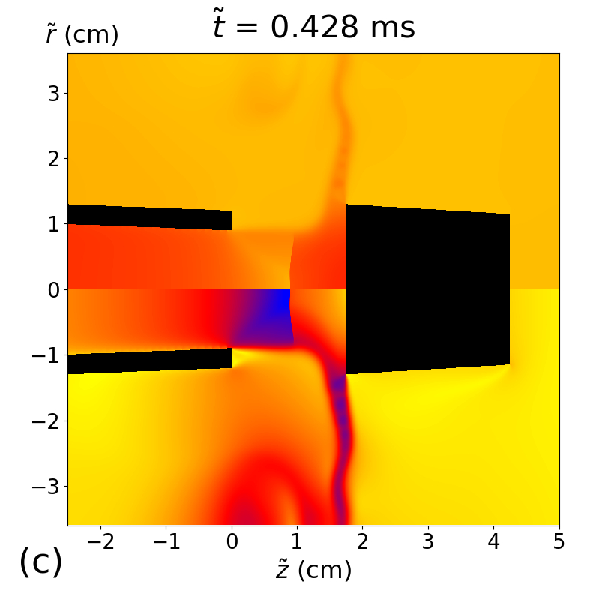}
	\includegraphics[width=0.512\textwidth]{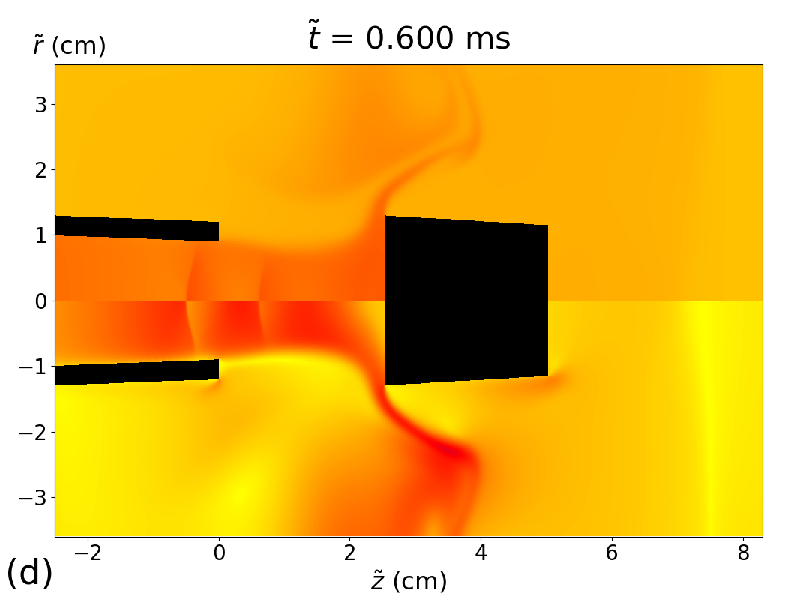}
	\caption{Temporal evolution of density (top halves of all graphs) and Mach number (bottom halves); two Mach discs are finally observed.}
    \vspace*{-1cm}
    \label{Kontur}
\end{figure}

The contour plots of the density (positive radii) and Mach number (negative radii) distributions shown in figure \ref{Kontur} for four different snapshots, taken from the supplementary movie ``Movie 2'', illustrate the evolution of the flow as their labels refer to the instances (a)--(d) itemised in the introduction of section \ref{s:rpd}. In particular, it becomes evident how the cylindrical stopper that has just escaped from the bottle is morphed into the terminal truncated cone. Its decompression causes an acoustic wave propagating in the radial direction (figure~\ref{Kontur}a). The Mach disc is generated at the jet edge roughly after $0.25$\,ms and subsequently moves in the positive $z$-direction (figure \ref{Kontur}b). This delay in its appearance at different values of $r$ explains the initial convex shape of the disc. Eventually, the disc reaches its maximum distance from the bottle opening at around $0.428$\,ms (figure \ref{Kontur}c), then moves in the opposite direction to retract into the bottle after about $0.56$\,ms (figure \ref{Kontur}d), thereby converted into a compression wave. In some cases it can even be reflected as a pressure wave at the liquid interface; see instance~(e).

In addition, a second, but very weak Mach disc split off the first one or created further upstream can only be observed near the axis, which is created during the retraction phase of the original disc (figure \ref{Kontur}d). The thin `boundary layers' visible in figures \ref{Kontur}b--\ref{Kontur}d are a purely numerical artefact resorting to the realisation of the no-penetration condition at a wall consisting of discrete steps. As another intriguing finding, figure \ref{Kontur} unveils that the jet becomes umbrella-shaped due to the impingement and deflection of the expanding gas at the rear side (base) of the cork (cf.\ the discussion of the waves in section \ref{ss:igs}). Thereby a number of supersonic `pockets' (blue colour regions) shaped as concentric tori temporally emerge (figures \ref{Kontur}a--c), in agreement with the findings of \cite{Beetal22}.

\begin{figure}
	\centering
    \hspace*{-4mm}\includegraphics[width=0.462\textwidth]{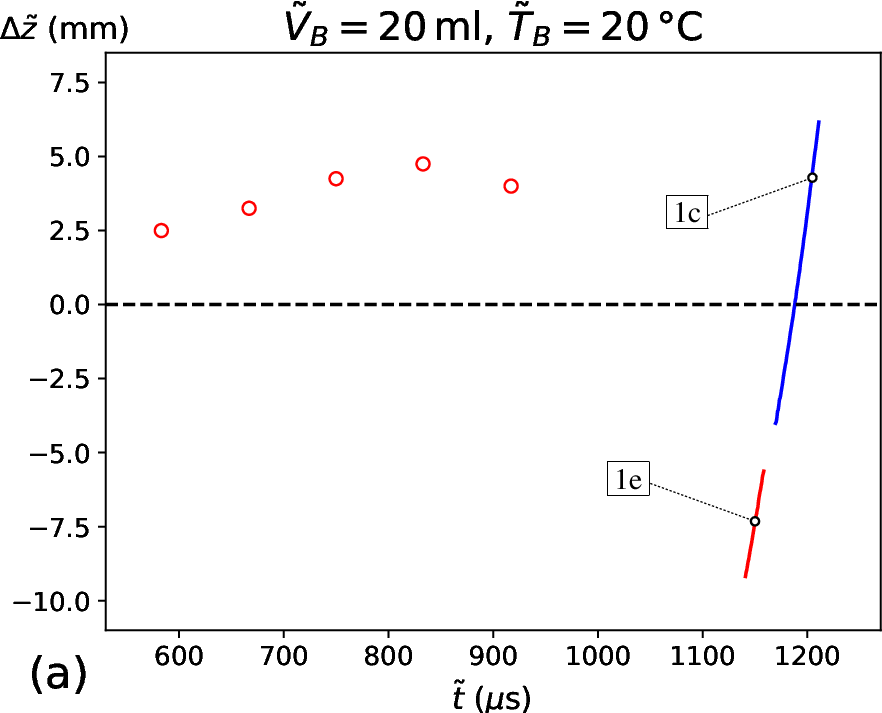}
    \vspace*{5mm}
	\hspace*{4mm}\includegraphics[width=0.445\textwidth]{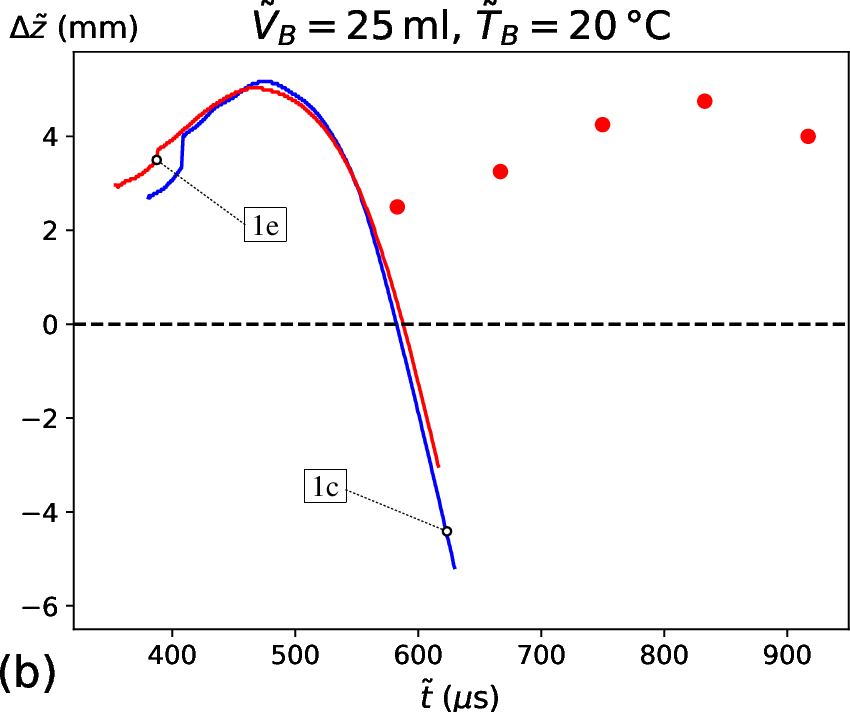}
	\hspace*{-1.7mm}\includegraphics[width=0.448\textwidth]{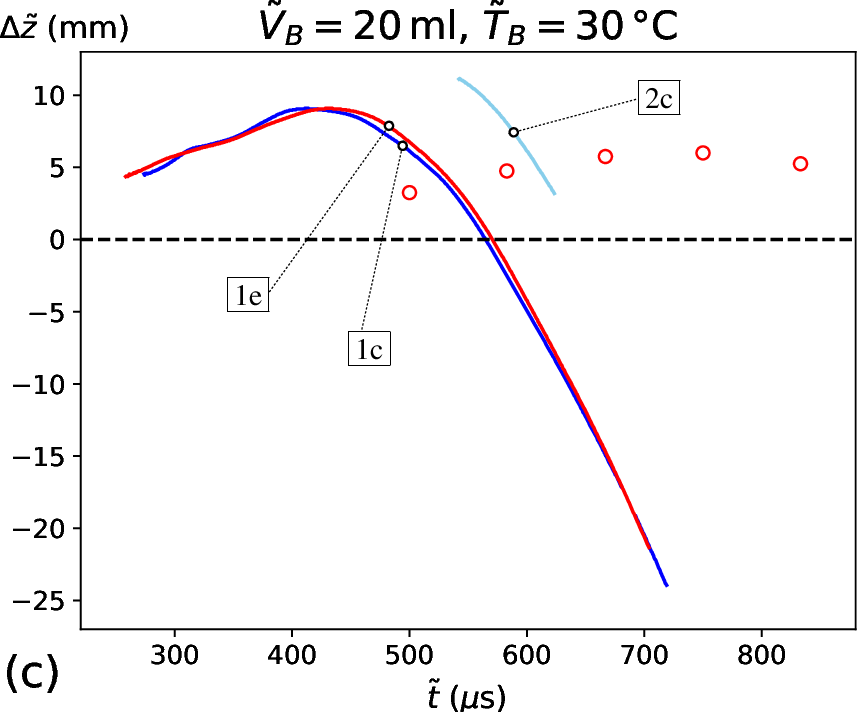}
    \hspace*{3.5mm}\includegraphics[width=0.45\textwidth]{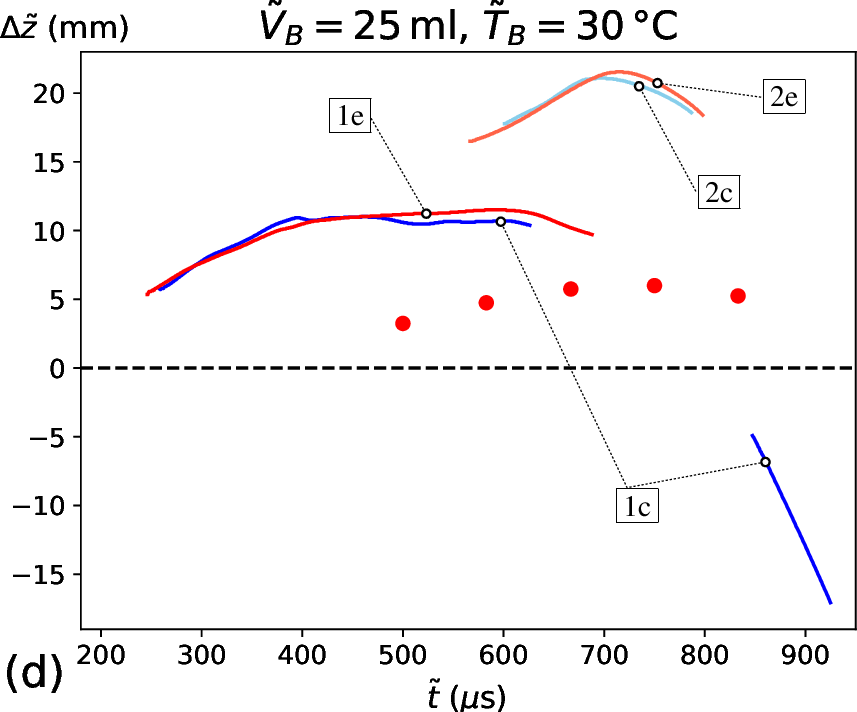}
	\caption{Distances $\Delta\tilde{z}$ of Mach disc ($\Delta\tilde{z}>0$) and compression wave ($\Delta\tilde{z}<0$) from bottle opening: numerical results (lines) for centre (blue) 
         and edge of the disc (red) vs.\ experimental ones for edge (red circles, for $\tilde{V}_B=25\,$ml from \citealp{Lietal19}, repeated for $\tilde{V}_B=20\,$ml); the cases (a)--(d) refer to their counterparts (A)--(D) in table \ref{t:cases}; darker lines refer to first and lighter ones to second Mach disc, only recorded in cases (C) and (D).}
        \label{VT_para}
\end{figure}

Figures \ref{VT_para}a--d visualise the quantitative behaviour of the Mach disc and the compression wave in terms of its axial distance from the bottle orifice, $\Delta\tilde{z}$, for the respective cases (A)--(D) in table \ref{t:cases}. Specifically, the discussion of figure \ref{Kontur} for the reference case (C) is condensed in figure \ref{VT_para}c.
According to the overview introducing section \ref{s:rpd}, the cases (b)--(d) show the behaviour of the first Mach disc, (c) and (d) additionally that of the second one, whereas the rather pathological case (a) refers to the end phase, (e): here our detection algorithm only records a Mach disc travelling outwards from the bottle at supersonic speed as accelerated by the still out-flowing gas. It vanishes after having reached its maximum distance. We admit that the choice of the numerical threshold entering the algorithm might also be relevant when it comes to the detected reappearance of the first Mach disc in case~(d).
 
While the distance between the two discs stays fairly constant, the life span of the second one is much shorter (figure \ref{VT_para}c). As expected, higher temperatures and thus pressures in the bottle (cf.\ table \ref{t:tpinp}) promote the dynamics of the Mach disc, in particular  $\Delta\tilde{z}_{\max}=\max(\Delta\tilde{z})$. This represents a definite reference quantity as it allows for, at least qualitatively, a comparison with the experimental data from \cite{Lietal19}. Increasing the internal volume $\tilde{V}_B$ and temperature $\tilde{T}_B\,(\propto\tilde{p}_B)$ of the bottle leads to a bigger value of $\tilde{F}_b$, which in turn increases the respective exit speed of the stopper, $\tilde{\Dot{Z}}$ at $\tilde{t}=0$, cf.\ figure \ref{v,Z}. Table~\ref{t:VT_para} shows these speeds together with the associated values of $\Delta\tilde{z}_{\max}$ and the time needed to reach it, $\Delta\tilde t_{\max}$, in qualitative agreement with our expectations and the experiments. This prompts us to conclude that $\Delta\tilde{z}_{\max}$ is (roughly) proportional to its speed. A similar trend can be found by analysing the curvature of the first Mach disc in figure \ref{VT_para}, which generally seems to be less for a faster stopper.

Surprisingly, as inferred from figure \ref{VT_para}, the simulated location of the Mach disc always occurs a certain  (constant) amount of time sooner than the corresponding sensed one. The definite clarification of this lag, perhaps being intrinsic to the method of sensing the disc edge, is a subject of future efforts.
\begin{table}
 \centering
 \begin{tabular}{cccc}
  \toprule
  \multicolumn{4}{c}{Case ($\#$): $\tilde{\Dot{Z}}$ (m/s), $\Delta\tilde{z}_{\max}$ (mm), $\Delta\tilde{t}_{\max}$ ($\mu$s)} \\[2pt]
  \midrule
  (A): $18.3$, $-5.6$, $1158$ & (B): $23.2$, $5.0$, $461$ & (C): $36.0$, $9.1$, $428$ & (D): $39.7$, $11.5$, $582$ \\[2pt]
  \bottomrule
 \end{tabular} 
 \bigskip
 \caption{Exit speed of the stopper and maximum disc distance from its edge for the cases in table~\ref{t:cases}.}
 \label{t:VT_para}
\end{table}

\vspace*{-5pt}
\section{Summary and further outlook}\label{s:sfo}

The present study resolves the axisymmetric complex gas dynamics accompanying the opening of a champagne bottle (or, speaking more generally, a bottle containing a pressurised liquid and gas) and its interaction with the propelled bottle stopper. For this purpose, the Euler equations were solved numerically via Godunov's method adopting a problem-tailored Roe solver (open source environment \textit{Clawpack}) and the no-penetration conditions, satisfied on the bottle and the stopper, and the outflow conditions implemented via ghost cells. Whilst the fluid-structure interaction results in a net pressure force accelerating the stopper in the axial direction, the sliding friction slows down the compressed stopper as long as it has not entirely passed the opening of the bottle. The hereby required surface stress is modelled via the typical hyperelastic constitutive law for cork, with the material parameters found by in-house experiments. After the stopper has fully escaped from the bottle, its radial expansion, modelled by elastic-wave propagation, causes its geometry to eventually attain the original, uncompressed truncated cone. The diameter of the bottle opening as well as the initial temperatures and pressures of the pressurised gas contained in the bottle agree with the values used by \cite{Lietal19}, which allows for a satisfactorily good agreement between our numerical and their experimental predictions.

The difference between the (scalar) pressure force at the base of the stopper, $\tilde F_b$, and the (scalar) bottle force, $\tilde F_B$, decisively controls its motion as the pressure forces acting on its remaining surfaces have negligible impact. Since $\tilde F_B$ decreases slower over time than $\tilde F_b$, the acceleration of the stopper must exhibit a minimum, found to be near $0.5\,$ms prior to its full escape. Even more, this minimum value can become so negative such that the stopper gets stuck inside the bottleneck. We adopted different strategies to asses the spatial and temporal consistency of the numerical method with respect to refined grid resolutions and maximum time steps.

As of utmost interest, a Mach disc forms between the bottle opening and the freely moving stopper. This initially exhibits a convex shape due to the radially varying times of the shock generation. The Mach disc reaches a maximum distance from the bottle opening and then retracts towards the latter. Moreover, during this phase a second disc is potentially created upstream of or split off the first. That maximum distance depends strongly on the input parameters, whilst the time between the full release of the stopper and its occurrence seems to only depend weakly on the initial conditions. Despite the good qualitative agreement between its simulated and experimentally found values, the nearly invariant offset between the measured and numerically predicted time of its emergence deserves to be unravelled. In future, evaluating the full set of Rankine--Hugoniot conditions might refine the shock detection. The following improvements may further help to reduce the deviation of the simulated from the measured data.

For instance, dealing with at least two species of gases, i.e.\ air and CO$_2$, is indicated, where Henry's law could provide the initial solubility of CO$_2$. Then the pressurised gas and the ambient air are more realistically treated as two different ideal gases. In an inviscid flow model, the thereby entailed immiscible two-phase flow would give rise to phase discontinuities. In turn, numerical dissipation would smooth them as it currently counteracts the resolution of the shocks and provokes the formation of artificial boundary layers and finally vortices at the later stages of the simulations. The viscous influence on the temporal approach towards equilibrium is assessed by balancing the instantaneous acceleration and the shear stress exerted across the diameter of the bottle opening: this gives a representative but irrelevantly long relaxation time $\tilde{d}_0^2\tilde{\rho}_0/\tilde{\eta}\doteq 21\,$s. The action of viscosity, however, is effectively at play and originates in the very first stages of the gap forming between the bottle opening and the stopper and its motion relative to the flow during later times. This is where the fully inviscid treatment apparently fails and calls for a more sophisticated analysis of these localised effects, involved by the largeness of the characteristic Reynolds number. We believe that such an approach, combined with solving the full Navier--Stokes equations, also in related situations will outweigh the straightforward use of turbulence models \citep{Beetal22}. Also, the shape and the material behaviour of the cork stopper could be modelled more realistically; the first by including its mushroom-like top, the latter by inclusion of wetting, an orthotropic material behavior and viscoelasticity, this for enabling an improved (rigorous) description of its expansion during its passage of the opening.
A simulation accounting for the fully three-dimensional flow for extended times and breaking the axial symmetry by small disturbances would presumably provide the typical period of time where stipulating that symmetry remains permissible.

We deliberately adopted the, without doubt unusually high, reference temperatures of $20$ and $30$°C provided by \cite{Lietal19} to validate our results. With that said, further progress demands pressure measurements in a sealed bottle for much lower, viable temperatures, ranging from $5$--$10$°C. Last but not least, the present findings suggest that sensing the position of the Mach disc provides, quite remarkably, an option to determine either the gas pressure or temperature inside a champagne bottle.

\begin{Backmatter}
\paragraph{Acknowledgements}
We gratefully acknowledge the setup of the experimental input and technical support (setup of tribometer) by Mr.\ T. Lebersorger and Mr.\ Ch.\ Haslehner (both AC2T research GmbH) and the referees' fruitful comments and suggestions.
\paragraph{Funding Statement}
L.W. and B.S. gratefully acknowledge partial funding by the Austrian Research Promotion Agency (project COMET-K2 \emph{InTribology}, FFG-No.\ 872176, project coordinator: AC2T research GmbH, Austria). The authors acknowledge TU Wien Bibliothek for financial support through its Open Access Funding Programme.
\paragraph{Declaration of Interests}
The authors declare no conflict of interest.
\paragraph{Author Contributions}
All authors created the research plan and the organisation of the paper and contributed equally to the writing of the manuscript, except of \S\,\ref{s:mi} and \S\,\ref{s:pmp}, both essentially formulated by B.S. The evaluation of the experimental data, the numerical calculations and the preparation of the figures and supplementary material were solely carried out by L.W.
%
%
\paragraph{Ethical Standards}
The research meets all ethical guidelines, including
adherence to the legal requirements of the study country.
\paragraph{Supplementary Material}
Supplementary information (technical details section, simulation movie) are available at 
\url{https://doi.org/10.1017/flo.2023.???}. Raw data and proprietary programming code are available from the corresponding author (L.W.).
\end{Backmatter}
\BeforeClearDocument{
\clearpage
%
\setcounter{equation}{0}
\setcounter{figure}{0}%
\setcounter{page}{1}%
\setcounter{section}{0}%
\renewcommand{\theequation}{S\,\arabic{equation}}%
\renewcommand{\thefigure}{S\,\arabic{figure}}%
\renewcommand{\thepage}{S-\arabic{page}}%
\renewcommand{\thesection}{Supplement {\Alph{section}}}%
\renewcommand{\thesubsection}{{\Alph{section}}.\arabic{subsection}}%
\renewcommand{\thesubsubsection}{\thesubsection.\arabic{subsubsection}}%
\pagestyle{myheadings}%

\section*{\Large Supplementary technical details}\label{s:S}
\bigskip
In this supplement, we provide the reader with a detailed description of how we closed the axial stress exerted by the tapered bottleneck and thus evaluated the equation of motion \eqref{ddZ}. If not indicated otherwise, cross references refer to the main document.

\section{Constitutive law and sliding friction: model and experimental validation}\label{S-s:cls}

According to the slenderness and the constitutive behaviour of the stopper anticipated in items (iii) and (iv) of \S\,\ref{ss:ba} and following \cite{Sa20} (see also \citealp{FePaAl14}), we express the typical nonlinear dependence of $\tilde{\sigma}$ on the relative compression $\varepsilon$ with sufficient accuracy as
\begin{equation}
 \tilde\sigma_n(\varepsilon)=\tilde A\varepsilon^3-\tilde B\varepsilon^2+\tilde C\varepsilon\,,\qquad
 \varepsilon(z,t)=1-R(z,t)/r_C\bigl(z-Z(t)\bigr)\,,
 \label{stress}
\end{equation}
The positive model coefficients $\tilde A$, $\tilde B$, $\tilde C$ must be determined empirically via a best fit. These together with a reliable estimate of a constant friction coefficient $\mu$ referring to the pairing cork/glass form the input parameters determining the mechanical behaviour of the cork. 

Both $\tilde{\sigma}_n(\varepsilon)$ and the friction coefficient $\mu$ were sensed with due rigour by in-house testings. To this end, we employed a reciprocating standard tribometer to apply $\tilde{\sigma}_n$ quasi-statically and $\tilde{\sigma}_n$ in an oscillatory fashion on appropriate cork samples. To ensure a constant normal compressive stress acting normal to the contact surface, the setup typically requires a relatively small, cuboid or cylindrical sample to accomplish the first task. Therefore, a cuboid with dimensions $\tilde{D}_1\times\tilde{D}_2\times\tilde{D}_3=14.0\times 12.5\times 17.5\,\mathrm{mm}^3$ was cut out of the mid of a stopper. Given the constitutive behaviour outlined in item (iv) in \S\,\ref{ss:ba}, here and below the indices refer to main axes of stress and strain. A normal compressive force $\tilde{F}_3$ was applied by a lever with a length ratio of one-to-five and a corresponding amount of weight hanged onto its other end. This yielded an indentation $\tilde{\delta}_3\;(>0)$ measured via a laser beam and indeed no perceptible transverse expansion. A controlled stepwise increase of $\tilde{F}_3$ allowed for measuring the stress $\tilde{\sigma}_3=\tilde{F}_3/(\tilde{D}_1\tilde{D}_2)$ as a function of the relative compression $\varepsilon_3=\tilde{\delta}_3/\tilde{D}_3$. The observed uniaxial states of stress and strain confirm the stipulated constitutive properties of cork. With these in mind, a permutation of the axes is confidently supposed to leave the results essentially unchanged, and the measured quantities $\tilde{\sigma}_3$ and $\varepsilon_3$ can be identified with respectively $\tilde{\sigma}_n$ and $\varepsilon$ for the axisymmetric case in \eqref{stress}. Hence, a least-square fit entailed $\tilde A=30.608\,$MPa, $\tilde B=11.982\,$MPa, $\tilde C=4.200\,$MPa (rounded). The so recorded relationship $\tilde{\sigma}_n(\varepsilon)$ is depicted in figure~\ref{Exp_Data}a.

The marked temporal increase $\tilde{\delta}_3$ at a constant load $\tilde{F}_3$ and correspondingly delayed approach of its terminal value, indicating the anticipated static dependence of $\tilde{\sigma}_3$ on $\varepsilon_3$, deserves mentioning. This creeping behaviour points to the inherent, but neglected, viscoelastic properties of cork. Even after approximately two minutes since the load had been applied, the stopper had not reached its final state yet. Likewise, a noticeable relaxation was observed when the load was released.

The dynamic measurements of $\mu$ were  carried out with a real stopper sliding over a glass surface in a straightforward fashion. They confirm the suggestion by \cite{FaFo98} that $\mu$ decreases slightly with increasing sliding speed. An analogous but more pronounced trend was revealed for increasing loads, $\tilde{F}_n$, applied normal onto the stopper and the glass plate underneath. Since the setup required the associated normal stress being definitely lower than $\tilde{\sigma}_n$ during the sliding of the stopper through the bottleneck, the found data shown in figure \ref{Exp_Data}b were extrapolated to suggest a nearly constant value of $\mu$ close to $0.2$. This is the figure we used initially in our simulations. We emphasise that published measurements of $\mu$ are scarce, and the work by \cite{FaFo98} represents an appreciated exception. Their experiments suggest $\mu$ being a factor $2$ to $3$ times higher, but the cork as well as the testing procedure they adopted differ from ours. Also, these authors mention the difficulty of obtaining reliable results in a systematic manner given the strong scatter of the data.  
\begin{figure}
	\centering
	\includegraphics[width=0.49\textwidth]{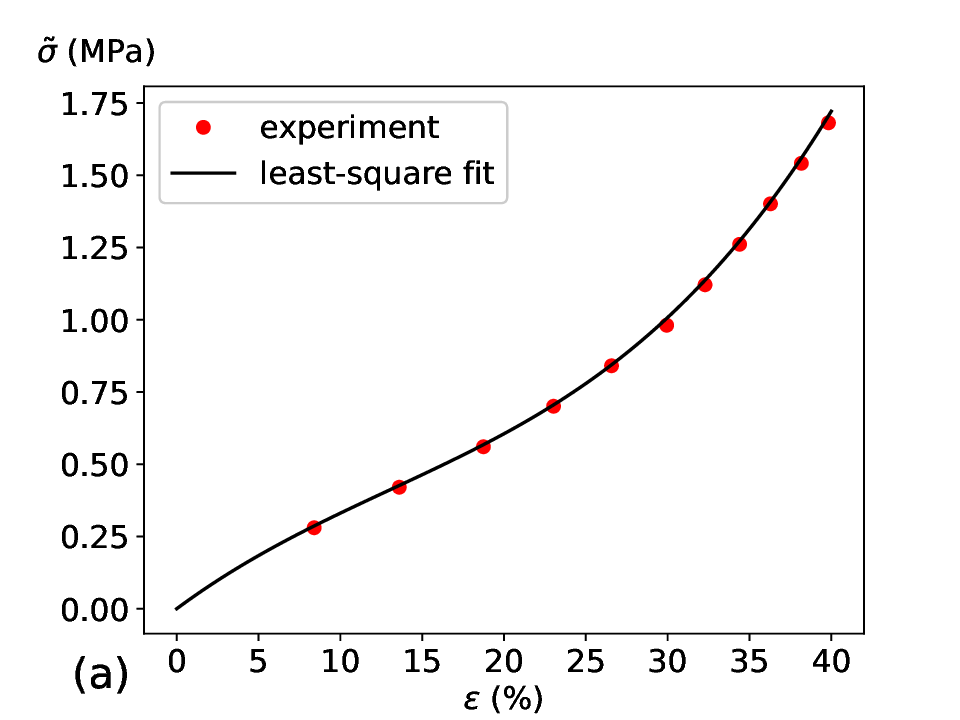}
	\includegraphics[width=0.49\textwidth]{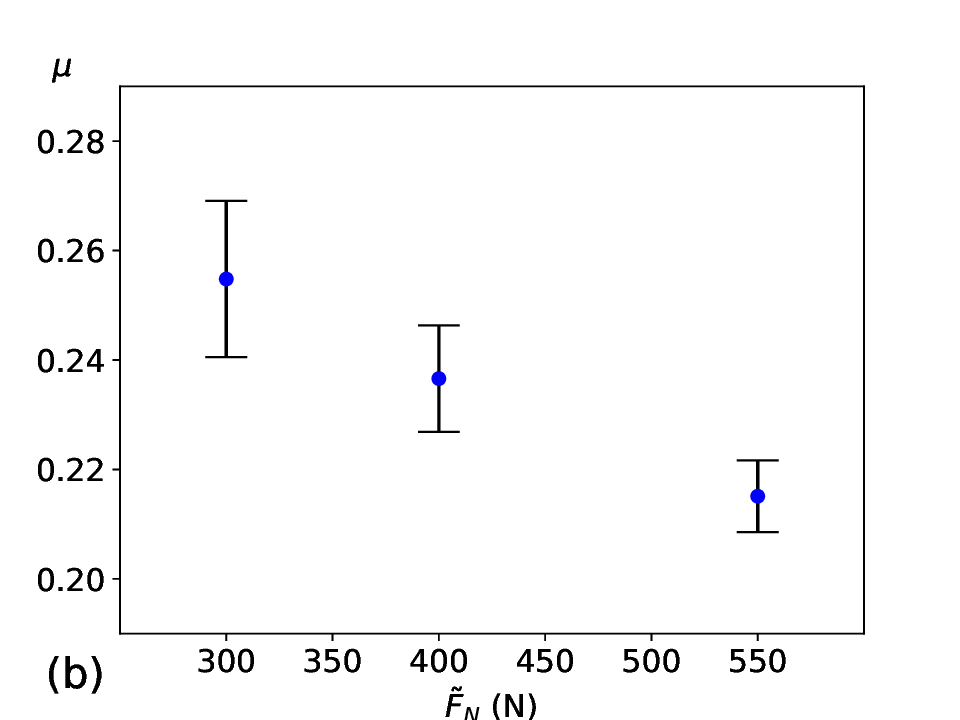}
	\caption{
        (a) Stress $\tilde{\sigma}_n$ exerted perpendicularly on stopper vs.\ its relative compression $\varepsilon$, \eqref{stress};
        (b)~friction coefficient $\mu$ by in-house experiments: mean values (blue dots) and observed variation (error bars).}
	\label{Exp_Data}
\end{figure}

\section{Axial reaction and friction force during sliding}\label{S-s:arf}

In this section, we scrutinise the numerical evaluation of \eqref{Fls_eq} and \eqref{FB}. It is expedient to change the variable of integration in \eqref{Fls_eq} to $r_C$ via $z=(d_2/2-r_C)/a_C+Z(t)$ and $\di z=-\di r_C/a_C$, see \eqref{rbrc}. This gives fixed lower and upper boundaries of integration $r_C=d_1/2$ and $d_2/2$, respectively. The still missing evaluation of $\partial R/\partial z$ is deferred to \ref{S-s:mse}.

Anticipating the constant slope $a_B=-\di r_B/\di z$, the above procedure of transforming the integration variable also applies to \eqref{FB}.
Expressing $\varepsilon$ likewise, we obtain
\begin{gather}
    F_B(t)=2\pi\,\frac{a_B+\mu}{a_C}\int\nolimits^{r_2}_{r_1(t)} \bigl[b_0\,r_C+b(t)\bigr]\;\sigma_n\Bigl(1-b_0-\dfrac{b(t)}{r_C}\Bigr)\di r_C\,, 
    \label{FBc}\\[3pt]
    b_0=a_B/a_C\,,\quad b(t)=1/2-a_B\bigl[Z(t)+r_2/a_C\bigr]\,,\quad r_1(t)=r_C\bigl(-Z(t)\bigr)\,,\quad r_2=d_2/2\,.
\end{gather}
The ratio $\mu/a_B\doteq 5.11$ for $\mu=0.2$ reveals a dominance of the frictional over the compressive contribution to $F_B$. We can express $F_B$ in closed form upon substitution of \eqref{stress} in \eqref{FBc} and integration of the resulting polynomial in $r_C$. This yields
\begin{equation}
    \begin{split}
     F_B(t)=2\pi\,\frac{a_B+\mu}{a_C}\,\Bigl(& \dfrac{a_0}{2}\bigl[r_2^2-r_1^2(t)\bigr]\,+a_1\, b(t)\bigl[r_2-r_1(t)\bigr]+a_2\,b^2(t)\ln\Bigl[\frac{r_2}{r_1(t)}\Bigr] \\[3pt]
     & -a_3\,b^3(t)\,\bigl[r_2^{-1}-r_1^{-1}(t)\bigr]+a_4\,b^4(t)\,\bigl[r_2^{-2}-r_1^{-2}(t)\bigr]\Bigr)\,,
    \end{split}
    \label{FBt}
\end{equation}
with the coefficients $a_{0,1,2,3,4}$ given by
\begin{equation}
    \left.
    \begin{aligned}
        a_0 &= b_0\bigl[A(1-b_0)^3-B(1-b_0)^2+C(1-b_0)\bigr]\,, \\
        a_1 &= A(1-b_0)^2(1-4b_0)-B(1-b_0)(1-3b_0)+C(1-2b_0)\,, \\
        a_2 &= -3A(1-b_0)(1-2b_0)+B(2-3b_0)-C\,, \\
        a_3 &= A(3-4b_0)-B\,,\qquad a_4=A/2\,.
    \end{aligned}\quad\right\}
\end{equation}

\section{Modelling the stopper's expansion}\label{S-s:mse}

In the light of its constitutive behaviour, the radial decompression of the stopper occurs in a nonlinear fashion. The two main assumptions underlying its modelling, anticipated in item (v) in \S\,\ref{ss:ba}, are: (I)~it starts not earlier as at $t=0$; (II)~it propagates with a speed comparable to that of a hyperelastic (radial-longitudinal) wave, i.e.\ given by $(\tilde{E}/\tilde{\rho}_C)^{1/2}$ where $\tilde{E}=\di\tilde{\sigma}_n/\di\varepsilon$ is the local bulk modulus of cork and $\tilde{\rho}_C$ its local density, until it has reached its relaxed state. We thus arrive at the modelled dimensionless expansion speed
\begin{equation}
    \Dot{R}(z,t)=\dfrac{\partial R}{\partial t}=
    \sqrt{\dfrac{1}{\kappa\,\rho_C(z,t)}\,\dfrac{\di\sigma_n\big(\varepsilon(z,t)\big)}{\di\varepsilon}}\,,\quad 
    \rho_C(z,t)=\rho_C^m\,\bigg[\dfrac{r_C\bigl(z-Z(t)\bigr)}{R(z,t)}\bigg]^2
    \label{se}
\end{equation}
under the constraints $R\leq r_C$, $Z(t)\leq z\leq Z(t)+l_C$, $t\geq 0$. Herein, $\rho_C=\tilde{\rho}_C/\tilde{\rho}_0$ and $\rho_C^m=\tilde{\rho}_C^m/\tilde{\rho}_0$ (see \S\;\ref{ss:gbu}). In order to initiate the expansion in a most smooth manner for the sake of numerical stability, the just escaped stopper is conveniently taken as cylindrical:
\begin{equation}
    R(z,t\leq 0)=
    \begin{cases}
        \;r_B(z) &\:\; \bigl(Z(t)\leq z\leq 0\bigr), \\[2pt]
        \;1/2 &\;\; \bigl(0<z\leq Z(t)+l_C\bigr).
    \end{cases}
\end{equation}
For $t=Z(t)=0$, we arrive at the initial condition $R(z,t=0)=1/2$. This complements \eqref{se} to a first-order problem governing the temporal variation of $R$, which is independent of the position of the stopper described by $Z(t)$ and thus decouples from the gas--stopper interaction. Therefore, the radial expansion solely depends on the geometry of the bottleneck apart from the material properties involved. 

Given the lossless elastic process it describes, \eqref{se} raises an oscillatory behaviour. In reality \citep{Lietal19}, the wavy surface of the stopper associated with its axial movement through the opening is rapidly damped by viscoelastic effects. Despite their entire neglect (last but not least, due to the lack of reliable data), the self-consistency of our simplified model, based on the premises (I) and (II) above, suggests a sufficiently reliable description of the expansion process. In order to suppress unphysically exaggerated oscillations, it is stopped for each value of $z$ at the instance of $t$ where $R$ reaches $r_C$ such that $R=r_C$ for all later times. The relaxation is taken as completed when $R=r_C$ for all values of $z$ along the stopper. We term the time needed to achieve this state relaxation time. Finally, $F_b$, $F_t$ and $F_{ls}$ in \eqref{Fbt} and \eqref{Fls_eq} can be evaluated.  

 Standard numerical integration of $\Dot{R}$ yields a reliable, universal value of approximately $40\,\mu$s for the relaxation time. The expansion of the base and of the top faces of the stopper predicted by our model is visualised in figure~\ref{vc,R}.
\begin{figure}
        \centering
 \hspace*{-0.3cm}\includegraphics[width=0.45\textwidth]{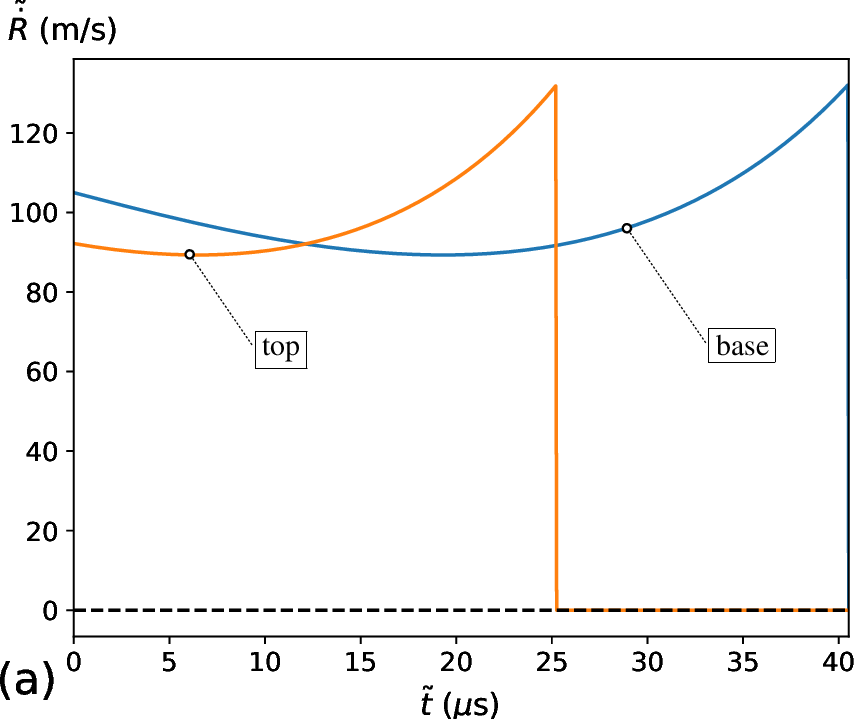}
	\hspace*{0.5cm}\includegraphics[width=0.45\textwidth]{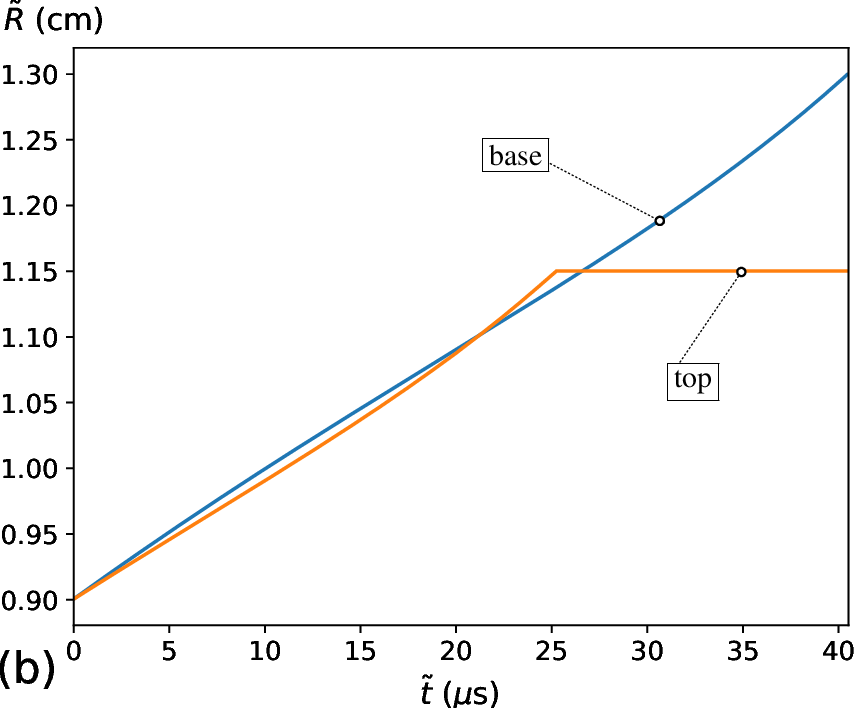}
	\caption{Expansion of the stopper: 
        (a)~radial speed of base (blue) and top surface (orange); 
        (b)~radii of base (blue) and top (orange) vs.\ $\tilde t$.}
	\label{vc,R}
\end{figure}

\section{Technical details of discretisation}\label{S-s:tdd}

Here we provide the interested reader with specific information concerning the numerical discretisation. Emphasis is placed on the dynamics of the stopper.

\subsection{Realisation of boundary conditions}\label{S-ss:GpBC}

At first, we want to support the understanding of the implementation of the boundary conditions, discussed in section \ref{s-BC}, with the graphical illustration in figure \ref{BC-fig}.
\begin{figure}[h]
	\centering
	\includegraphics[width=0.49\textwidth]{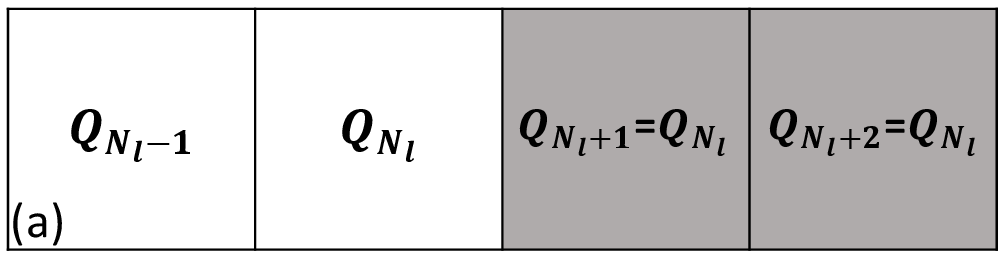}
	\includegraphics[width=0.49\textwidth]{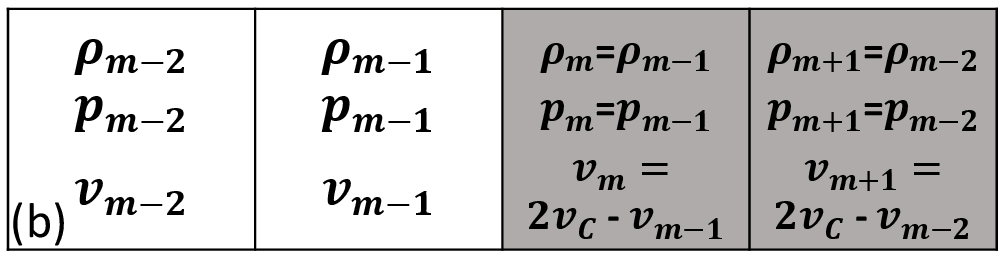}
	\caption{(a) Extrapolation condition and (b) no-penetration condition with a boundary velocity of $v_C$. Ghost cells in both figures are to the right and their total number is $N_G=2$. $N_l$ is the last cell inside the computational domain and $m$-$1/2$ is the surface index of the solid body.}
    \label{BC-fig}		
\end{figure}
\subsection{Gas--stopper interaction}

In order to assign the initial solution vector $\textbf{Q}^0$ to the correct position on the grid of the computational domain, functions must be defined which return the index of the cell interface nearest to a given point, denoted by the argument of an array inside square brackets. For a two-dimensional array the first index $i$ always refers to the $i$-th cell interface in the $z$-direction and the second one $j$ to the index in the $r$-direction, not depicting their potential dependence on time in any further equation. The opening of the bottle with index $b_o$ and the one for the base surface of the stopper $c_b(t)$ are evaluated by
\begin{equation}
    b_o\colon\hspace{3mm} \big|Z_i[b_0]\big|\rightarrow \min\,,\hspace{10mm}
    c_b(t)\colon\hspace{3mm} \big|Z_i[c_b(t)]-Z(t)\big|\rightarrow \min\,, 
    \label{b0cb}
\end{equation}
where $Z_i$ is the array containing all cell interfaces in the $z$-direction.
While $b_o$ and $c_b$ only possess one numerical value for each time step, the indices of the inner surface of the bottle $b_{is}$ and the ones for the lateral surface of the stopper $c_{ls}$ must be defined as an array with index $i$, including all cell interfaces with an index of $j$ and a position $R_i[j]$ in the $r$-direction for which the following conditions are fulfilled:
\begin{equation}
    \begin{split}
        b_{is}[i]\colon\hspace{3mm} \Big|R_i\big[b_{is}[i]\big]-r_B\big(Z_i[i]\big)\Big|\rightarrow\min\,,
        \hspace{20.5mm} i = 0,\dots,b_o-1\,,\\
        c_{ls}(t)[i]\colon\hspace{3mm} \Big|R_i\big[c_{ls}(t)[i]\big]-R\big(Z_i[c_b(t)+i],t\big)\Big|\rightarrow\min\,,
        \hspace{5mm} i = 0,\dots,c_l-1\,.
    \end{split}
    \label{bis,cls}
\end{equation}
Here $c_l=b_o-c_b(t_0)$ is the total number of cells occupied by the stopper in the $z$-direction, whereas $t_0$ is the starting time of the simulation. Here $b_o$ and $c_l$ are excluded because the last interface index corresponds to the first cell centre index outside the given object.

In order to fulfil the given boundary conditions, an indicator field, ind$[i,j]$, is laid over the computational domain where its values refer to the type of material: $0$ to fluid; $1$ to bottle; $2$ to stopper. Only if the corresponding values between two neighbouring cells are different, the boundary conditions are applied. The initial indicator field, ind$^0$, has the following form:
\begin{equation}
    \text{ind}^0\big[i,j\big] = 
    \begin{cases}
    \hspace{1mm}1\colon\hspace{2mm} i = 0,\dots,b_o-1\,,\hspace{15.5mm} 
    j = b_{is}[i],\dots,b_{os}[i]-1\,,\\
    \hspace{1mm}2\colon\hspace{2mm} i = c_b(t_0),\dots,c_t(t_0)-1\,,\;\;\, 
    \hspace{1mm}j = 0,\dots,c_{ls}(t_0)[i-c_b(t_0)]-1\,,\\
    \hspace{1mm}0\colon\hspace{2mm}\text{else}\,.
    \end{cases}
\end{equation}
Here $c_t(t)=c_b(t)+c_l$ is the index of the top interface of the stopper and $b_{os}[i]=b_{is}[i]+\Delta j_b$ are the indices of the outer surface of the bottle with a constant total number of glass cells $\Delta j_b$ in the $r$-direction.

The pressure forces on the base and top surface can now be discretised as
\begin{equation}
    F_b(t) \approx 2\pi\,\Delta r\hspace{-3mm}\sum_{j=0}^{c_{ls}(t)[0]-1}\hspace{-3mm}R_i[j]\;p(t)[c_b(t)-1,j] 
    \label{Fb_discret}\,,\quad
    F_t(t) \approx 2\pi\,\Delta r\hspace{-3mm}\sum_{j=0}^{c_{ls}(t)[-1]-1}\hspace{-3mm}R_i[j]\;p(t)[c_t(t),j]\,,
\end{equation}
where $i=-1$ indicates the last entry, for instance $c_{ls}(t)[\!-\!1] = c_{ls}(t)[c_l-1]$. Due to the expansion of the stopper being faster than its axial motion, for a short period of time $c_{ls}(t>0)[0]$ can be bigger than $b_{os}[\!-\!1]$, while $c_b(t>0)$ is still equal to $b_o$. Therefore the gas still trapped inside the bottle and the ambient air can both contribute to $F_b$, splitting the sum in (\ref{Fb_discret}.1) into
\begin{equation}
    \sum_{j} \rightarrow \sum_{j_1} + \sum_{j_2}\,\colon\quad j_1 = 0,\dots,b_{is}[\!-\!1]-1\,,
    \quad j_2 = b_{os}[\!-\!1],\dots,c_{ls}(t)[0]-1\,.
\end{equation}
While $F_b$ and $F_t$ are integrated over $r$, the pressure force $F_{ls}$ acting on the lateral surface of the stopper must be integrated over $z$. This is referenced by the first sum in
\begin{equation}
    F_{ls}(t\geq0)\approx\,2\pi\,\Delta r \sum_{i\in \,I_{ls}(t)} {\sum_{j}}^\pm \hspace{0.5mm} R_i[j]\hspace{1mm}p(t)[c_b(t)+i,j]\,. \label{Fls_discret}
\end{equation}
\begin{figure}
	\centering
	\includegraphics[width=0.49\textwidth]{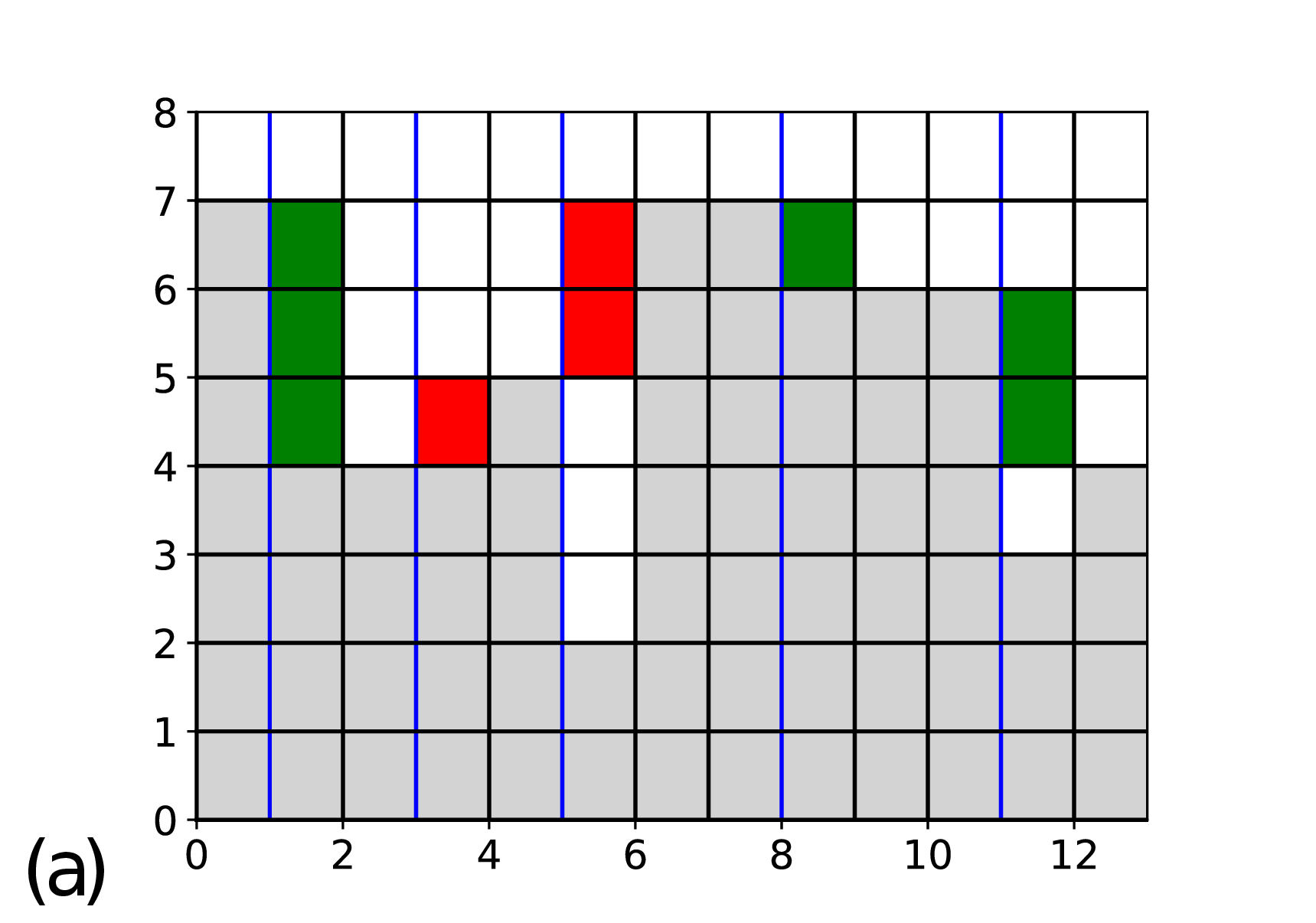}
	\includegraphics[width=0.49\textwidth]{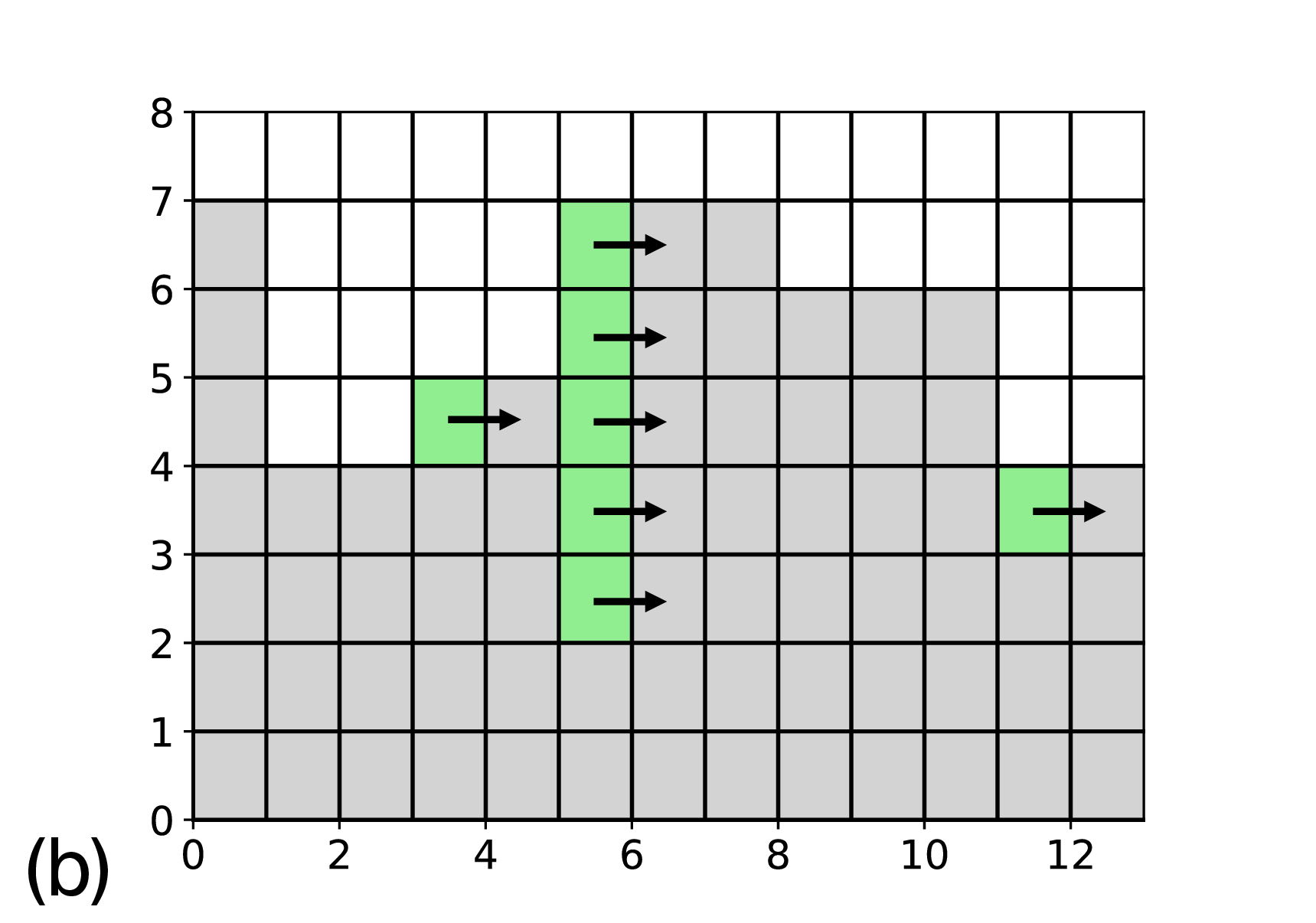}
	\caption{(a) Fluid cells contributing to $F_{ls}$: the stopper occupies the grey cells, the fluid the remaining ones; the red cells reduce and the green ones increase $F_{ls}$; the blue lines indicate the values of $I_{ls}=\{1,3,5,8,11\}$. 
	(b)~Shifted fluid cells (light-green) due to axial movement of the stopper.}
    \label{Ils,shift}		
\end{figure}
The array $I_{ls}$ contains the indices relative to $c_b(t)$ of all fluid cells in the $z$-direction contributing to $F_{ls}$. For bigger surface gradients, multiple cells can possess the same entry in $I_{ls}$ (see figure~\ref{Ils,shift}a), explaining the second sum in \eqref{Fls_discret}, which is defined as
\begin{equation}
    {\sum_{j}}^\pm =
    \begin{cases}
         \hspace{2.8mm}\sum_{j=c_{ls}(t)[i+1]}^{c_{ls}(t)[i-1]-1}\colon & c_{ls}(t)[i-1]>c_{ls}(t)[i+1]\,,\\[3mm]
        -\sum_{j=c_{ls}(t)[i-1]}^{c_{ls}(t)[i+1]-1}\colon & c_{ls}(t)[i-1]<c_{ls}(t)[i+1]\,,
    \end{cases}\hspace{5mm} i \in I_{ls}(t)\colon\;\; 0 < i < c_l-1\,.
    \label{sum pm}
\end{equation}
This distinction is necessary to describe fluid cells acting on left (\ref{sum pm}, top row) and right (\ref{sum pm}, bottom row) interfaces describing the discrete lateral surface of the stopper. As an immediate consequence of its expansion, $I_{ls}$ and therefore also ${\sum_j}^\pm$ can change at every time step.

\subsection{Axial motion of the stopper}

The spatially continuous motion of the stopper for discrete but variable time steps was already shown in \eqref{Zn}. $t_n=0$ is thereby defined as the point in time, when $c_b(t_n) = b_o$ for the first time. 
The result in \eqref{Zn} is then used to calculate $c_b(t_{n+1})$ with \eqref{b0cb}, performing the following updates only if $c_b(t_{n+1})\neq c_b(t_n)$.
\begin{enumerate}
    \item Check if $c_b(t_{n+1})-c_b(t_n)=1$, otherwise the stopper has moved too fast and a smaller $\Delta t_n$ must be chosen.
    \item All values of $I_{ls}$ increase by one due to the stopper moving exactly one cell.
    \item The cells behind the stopper are now fluid cells,
    \begin{equation}
        \text{ind}[c_b(t_n),j]=0\,\colon\hspace{5mm} j =
        \begin{cases}
            \hspace{1mm}0,\dots,b_{is}[c_b(t_n)]-1\colon & t_n < 0\,,\\
            \hspace{1mm}0,\dots,c_{ls}(t_n)[0]-1\colon & t_n \geq 0\,,
        \end{cases}
    \end{equation}
    and are filled with the same values as the ones to their left:
    \begin{equation}
        \textbf{Q}[c_b(t_n),j] = \textbf{Q}[c_b(t_n)-1,j]\,.
    \end{equation}
    \item At that moment when there is only a single cell between the bottle opening and the stopper ($c_b(t_{n+1}) = b_o+1$) in the $z$-direction, the values of the cells in between these two objects are taken from the ones above the bottle:
    \begin{equation}
        \textbf{Q}\big[b_o,j\big] = \textbf{Q}\big[b_o-1,j+\Delta j_b\big]\,\colon\hspace{5mm} j = b_{is}[\!-\!1],\dots,b_{os}[\!-\!1]-1\,.
    \end{equation}
    \item Finally, the cells occupied by the stopper are updated in such a way that
\end{enumerate}
    \begin{equation}
        \begin{split}
            \textbf{Q}[c_b(t_{n+1})+i,j] = \mbox{NaN}\,,\\
            \text{ind}[c_b(t_{n+1})+i,j] = 2\,, 
        \end{split}\hspace{2mm}\Bigg\}\colon\hspace{2mm}
        i,\,j=\Bigg\{
        \begin{split}
            &c_l-1,\hspace{9mm} 0,\dots,b_{is}[-1]-1\colon  \hspace{6mm} t_{n} < 0\,,\\
            0,&\dots,c_l-1,\hspace{2mm} 0,\dots,c_{ls}(t_{n+1})[i]-1\colon 
            \hspace{2mm}t_{n}\geq 0\,.
        \end{split}
        \label{update_stop}
    \end{equation}

\begin{figure}
	\includegraphics[width=0.49\textwidth]{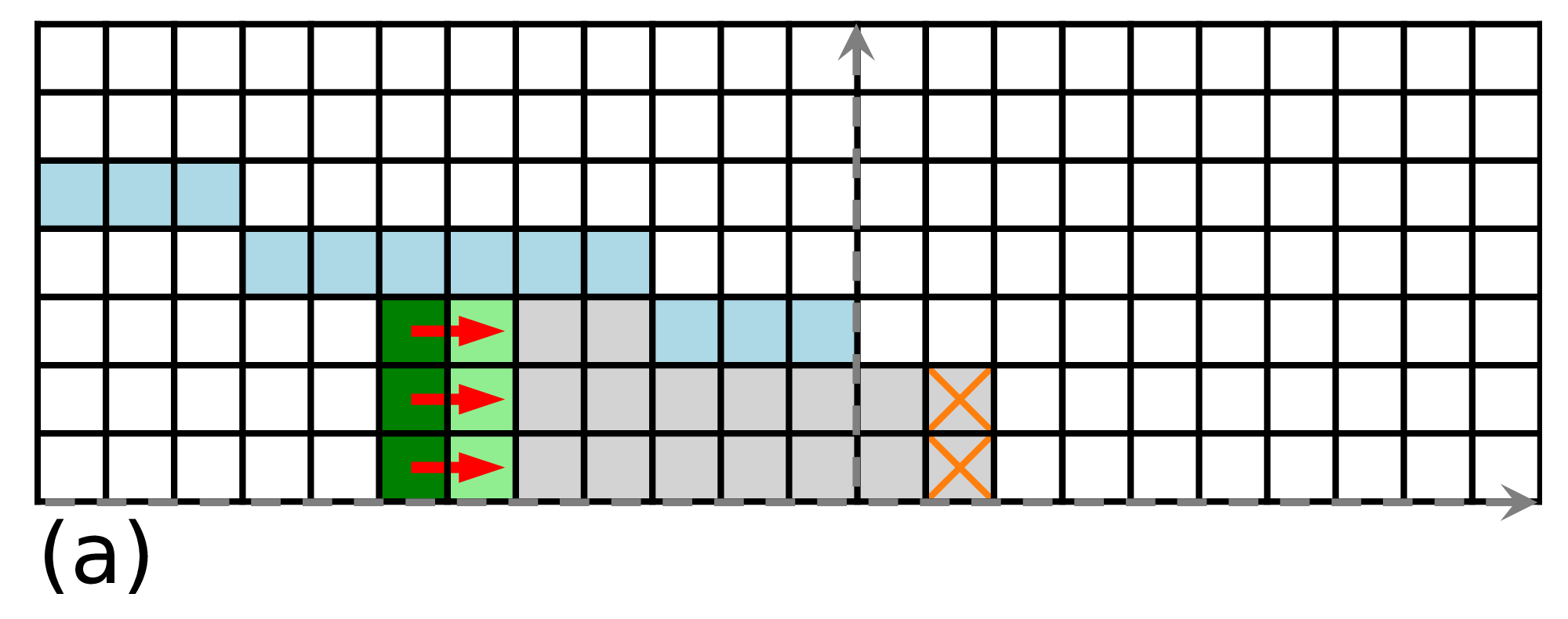}
	\includegraphics[width=0.49\textwidth]{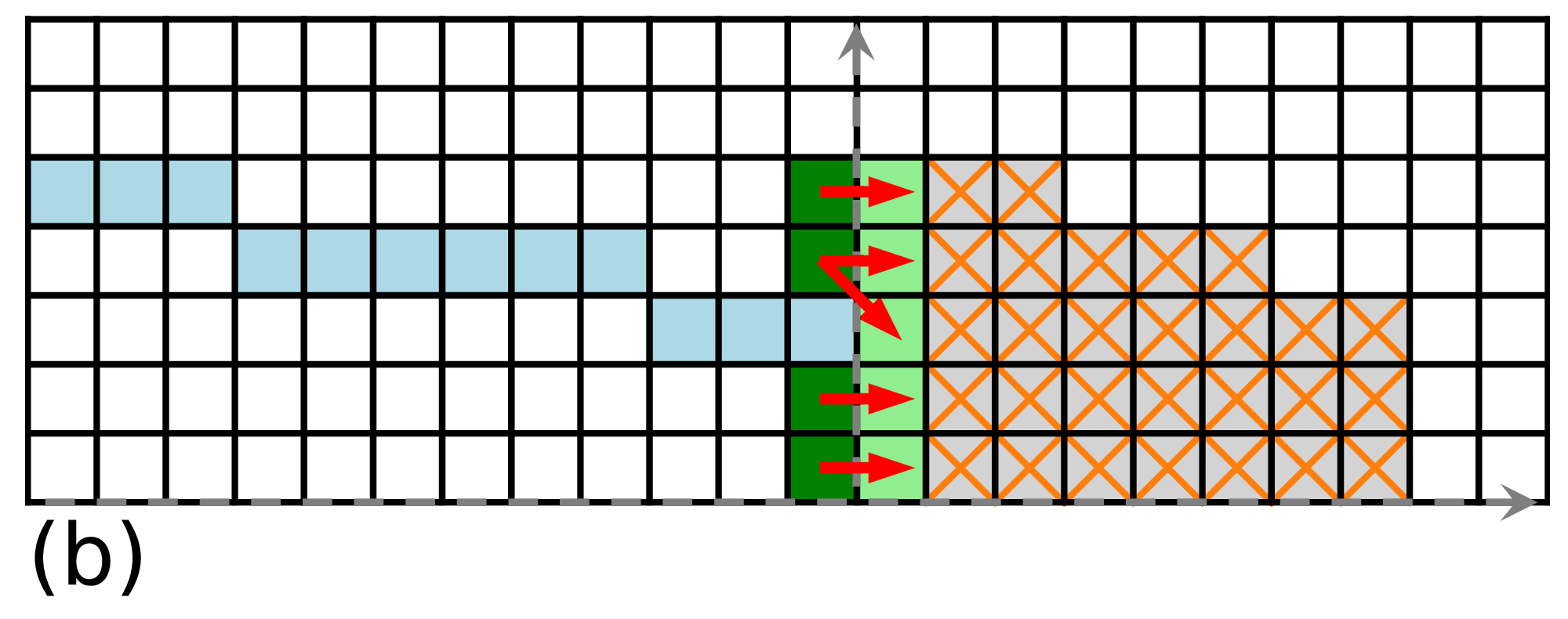}
	\caption{Cell updates for $t_{n}<0$ (a) and $t_{n}\geq0$ (b). The green fluid cells update the light green ones behind the stopper (grey). While the cells of the bottle (blue) keep their initial values, the cells of the stopper marked with orange crosses have to be changed accordingly (ind=$2$, $Q$=\textit{NaN}).} \label{Grid_tpos}	
\end{figure}
\subsection{Radial expansion of the stopper}

While the axial motion of the stopper is characterised by updating $c_b$, the indices of the lateral surface have to be computed prior to that because when assigning the new ind- and $\textbf{Q}$-values, \eqref{update_stop} already uses $c_{ls}$ for the new time frame $t_{n+1}$. The temporally discrete radius of the stopper $R^{n}[i]$ is spatially continuous with respect to $r$, but is defined as an array with index $i = 0,\dots,c_l-1$, describing its discrete nature with respect to $z$. The expansion speed $\Dot{R}^n$ is evaluated as mentioned in \eqref{se}, exchanging $R(z,t)\rightarrow R^n[i]$, $Z(t)\rightarrow Z^n$ and $z\rightarrow Z_i[c_b(t_n)+i]$. The latter two are the discrete arguments of $r_C$ ($Z_i-Z^n$). $R^{n+1}[i]$ is computed with the explicit Euler method. In a following step, $R^{n+1}[i]$ is used in (\ref{bis,cls}.2), finding the solution for $c_{ls}(t_{n+1})[i]$. If $c_{ls}(t_{n+1})\neq c_{ls}(t_{n})$ is true for at least one entry, $I_{ls}(t_{n+1})$ and $\sum_{j}^\pm$ are updated accordingly. 

Next, the algorithm for the axial motion of the stopper starts off, updating $\textbf{Q}$, $Z$ and $R$, the first two only if necessary. The only exception is the last step, described in \eqref{update_stop}, which must be applied even if the sole discrete motion of the stopper is its radial expansion for the given time frame, increasing the number of cells occupied by the stopper. If the object also or exclusively moves axially at that instant, the whole body is shifted along the $z$-axis, which requires proper allocation of the new fluid cells (see figure~\ref{Ils,shift}b):
\begin{equation}
 \begin{split}
    &\text{ind}[c_b(t_{n+1})+i,j]  =  0\,, \\
    \textbf{Q}[c_b(t&_{n+1})+i,j] = \textbf{Q}[c_b(t_n)+i,j]
 \end{split}\hspace{2mm}\Bigg\}\colon\hspace{2mm}
 \Bigg\{
 \begin{split}
    0 \leq i < c_l-1,\quad c_{ls}(t_n)[i] < c_{ls}(t_n)[i+1]\,,  \\
     j = c_{ls}(t_n)[i],\dots, c_{ls}(t_n)[i+1]-1\,.
 \end{split} 
 \label{Q,aux shift}
\end{equation}
\eqref{Q,aux shift} is only invoked during the expansion phase since fluid cells need no longer to be shifted along the lateral surface of the stopper after it has reached its relaxed state. Furthermore, it is important that the shift of indices happens before \eqref{update_stop}; otherwise, newly created solid cells are shifted instead.
}
\end{document}